\documentclass[lettersize,journal]{IEEEtran}
\usepackage{amsmath,amsfonts}
\usepackage{array}
\usepackage{textcomp}
\usepackage{stfloats}
\usepackage{url}
\usepackage{verbatim}
\usepackage{graphicx}
\usepackage{cite}

\usepackage{multirow}
\usepackage{subcaption}
\usepackage{booktabs}
\usepackage{color}
\usepackage[table,xcdraw]{xcolor}
\usepackage[vlined,ruled]{algorithm2e}
\usepackage{acronym}
\acrodef{ast}[AST]{Audio Spectrogram Transformer} 
\acrodef{amfm}[AM-FM]{Amplitude-and Frequency-Modulated}
\acrodef{drc}[DRC]{Dynamic Range Compression}
\acrodef{tfr}[TFR]{Time-Frequency Representation}
\acrodef{mfcc}[MFCC]{Mel Frequency Cepstral Coefficients}
\acrodef{mee}[MEE]{Music Effect Encoder}
\acrodef{amfm}[AM-FM]{Amplitude-and Frequency-Modulated}

\begin{document}

\title{Neural-Enhanced Dynamic Range Compression Inversion: A Hybrid Approach for Restoring Audio Dynamics}

\author{
\IEEEauthorblockN{
Haoran Sun,
Dominique Fourer,
Hichem Maaref
}
\\
\IEEEauthorblockA{
\textit{IBISC (EA4526), Univ. Evry - Paris-Saclay} \\
haoran.sun@etu-upsaclay.fr, dominique.fourer@univ-evry.fr, hichem.maaref@univ-evry.fr
}
}




\maketitle

\begin{abstract}
Dynamic Range Compression (DRC) is a widely used audio effect that adjusts signal dynamics for applications in music production, broadcasting, and speech processing.
Inverting DRC is of broad importance for restoring the original dynamics, enabling remixing, and enhancing the overall audio quality. Existing DRC inversion methods either overlook key parameters or rely on precise parameter values, which can be challenging to estimate accurately. To address this limitation, we introduce a hybrid approach that combines model-based DRC inversion with neural networks to achieve robust DRC parameter estimation and audio restoration simultaneously.
Our method uses tailored neural network architectures (classification and regression),
which are then integrated into a model-based inversion framework to reconstruct the original signal.
Experimental evaluations on various music and speech datasets confirm the effectiveness and robustness of our approach, outperforming several state-of-the-art techniques.
\end{abstract}


\section{Introduction}
\ac{drc} is a fundamental process in audio signal processing that aims at changing the dynamic range of a signal. This technique is widely used in various stages of audio production, such as recording, mixing, and mastering, to control the loudness of an audio signal and to prevent clipping or distortion \cite{zolzer2002dafx}.
However, the application of \ac{drc} can modify the audio's timbre and the perceived quality, making its inversion a challenging task. 
Thus, inverting \ac{drc} is of great interest in the context of audio reverse engineering \cite{barchiesi2010reverse} since it aims at recovering the original dynamic range and audio quality of a signal.
This task has numerous potential applications, including signal restoration, audio remixing, and enhanced creative control.

Inverting \ac{drc} is a challenging problem that often requires
side information with an explicit \ac{drc} model or prior knowledge about the \ac{drc} parameters to be efficiently processed.
There exist only a few studies that directly address the problem of \ac{drc} inversion.
In \cite{lachaise2008inverting}, the authors consider \ac{drc} inversion as a rate-distortion optimization problem using a coder-decoder framework that minimizes both the side information and the reconstruction error when combined with a specific estimator applied to the compressed signal.
In \cite{GorlowDC}, the authors propose a specific \ac{drc} model, which provides a promising reconstruction approximation but requires knowing exactly the \ac{drc} parameters of the compressed signal. Other work, such as \cite{patel2020acoustic}, only attempts to cancel or reduce the effects of \ac{drc} without directly addressing the challenging inversion problem.
More recent inversion methods based on deep learning only consider a specific type of \ac{drc} effect (e.g., limiter, distortion, clipping, etc.), which may require a specific model and training dataset for each of them \cite{de-limiter, rice2023general}.
In \cite{take2024audio}, the authors propose a general audio effects chain estimation and dry signal recovery model, but its accuracy can be limited due to its lack of specificity.


In the present work, we propose to revisit the \ac{drc} inversion problem by resorting to a deep neural network through a new approach, which first identifies the mixture configuration to obtain the \ac{drc} parameters and then applies the adequate model-based inversion to restore the original uncompressed signal.

For the first step, we propose a hybrid approach. On the one hand, a classification model is used to identify the DRC configuration and then find the exact DRC parameters when there is a choice. On the other hand, a regression model is used to estimate the DRC parameters directly when there is no prior condition. 
Although predicting the DRC profile can yield accurate DRC parameters, directly estimating the parameters is more general and not constrained by the specific type of DRC profile.

For the audio classification technique, early approaches predominantly use spectral features and statistical descriptors combined with classical machine learning algorithms \cite{svm, knn}. While these methods provide interpretable results, they often struggle with capturing complex audio patterns effectively and require manual feature engineering. In contrast, deep learning methods, particularly Convolutional Neural Networks (CNN) \cite{hershey2017cnn} and Recurrent Neural Networks (RNN) \cite{rnn}, have gained prominence due to their ability to learn hierarchical representations from raw audio waveforms or spectrograms automatically.
Among deep learning-based audio classification methods, the Audio Spectrogram Transformer \cite{AST} (AST) has emerged as a powerful approach for capturing both temporal and spectral features of audio signals. In view of the excellent performance of transformer-based models in audio classification tasks in existing studies, we chose this model in our present work.

For the parameter estimation technique, several studies have extended the use of early machine learning techniques to parameterize non-linearities in physical models \cite{drioli1998learning, uncini2002sound}. Another computational intelligence approach \cite{itoyama2014parameter} estimates synthesizer parameters using a multivariate linear regression model with hand-crafted features. More in line with our present study is \cite{zhou2024audio} that uses a U-Net model to estimate parameters of multiple audio effects with the help of raw audio. 
In \cite{beafx}, 3 different encoders are used for audio effect parameters estimation, where the \ac{mee} \cite{mee} performed the best for the compression task.
Being inspired, in this work, we use \ac{mee} for estimating the \ac{drc} parameters.

Our contributions are manifold.
First, we propose a novel technique based on a supervised deep neural network model that predicts the \ac{drc} parameters applied to an audio signal by analyzing the waveform of the compressed signal.
Second, we use the estimated parameters corresponding to the predicted profile with a model-based \ac{drc} inversion technique introduced in \cite{GorlowDC} to restore the original signal $x$.
Finally, we compare our proposed approach with several state-of-the-art methods applied to public music audio and speech datasets, in terms of parameter estimation accuracy and signal reconstruction quality.

Our paper is organized as follows. 
Section~\ref{sec:method} introduces the \ac{drc} inversion problem and describes our novel approach. 
Section~\ref{sec:exp} presents the experimental setup. 
Section~\ref{sec:opt} presents the algorithm optimization and robustness analysis.
Section~\ref{sec:results} presents the numerical results.
Finally, Section~\ref{sec:conc} contains our conclusion and future work directions.

\section{Background on Model-Based DRC Inversion}\label{sec:method}
\subsection{Problem Fomulation}\label{sec:problem}
Let $x \in \mathbb{R}^N$ be a one-dimensional real-valued discrete-time signal sampled at rate $F_s$ expressed in Hz.
We denote $x[n]$, $\forall n \in \{0,1,\cdots,N-1\}$ the sample at index $n$.

Now, we consider $g$, the gain function of the \ac{drc} effect applied to $x$, which yields the compressed signal:
\begin{equation}
y[n] = x[n] \cdot g_{x, q_{\theta}}[n]. \label{eq:drc}
\end{equation}
The gain value $g[n]$ depends on the signal $x$ itself and the \ac{drc} parameters $q_{\theta} \in \mathbb{R}^7$, where $\theta \in \{0, 1, ..., d-1\}$ denotes the compressor profile label (i.e., the DRC profile), with $d$ the maximum number of considered profiles.

Hence, this paper addresses the problem of blindly computing the estimates $\hat{q_{\theta}}$ and $\hat{x}$, which are as close as possible to the ground truth $q_{\theta}$ and $x$ (in the minimum mean-squared error sense).
We consider the compressed signal $y$ as a unique observation, and assume that the applied \ac{drc} effect can be approximated using the proposed generic \ac{drc} model for which the set of parameters $q$ is unknown.

\subsection{DRC Principle}\label{sec:DRC}

The reversible \ac{drc} model proposed in \cite{GorlowDC} enables to consider a large variety of compressor types (eg. expander, compressor, or noise gate, etc.) and uses the following set of parameters (cf. Table~\ref{tab:5DRC} for a description) $q_{\theta}=\{L, R, \tau_v^{\text{att}}, \tau_v^{\text{rel}}, \tau_g^{\text{att}}, \tau_g^{\text{rel}}, p\}$, where
\begin{itemize}
 \item $L$: the threshold expressed in dB,
 \item $R$: the compression ratio,
 \item $\tau_v^{\text{att}}$, $\tau_v^{\text{rel}}$: the attack and release time (expressed in seconds), used to smooth the detection envelope,
 \item $\tau_g^{\text{att}}$, $\tau_g^{\text{rel}}$: the attack and release time (expressed in seconds), used to smooth the gain function,
 \item $p$: the compressor detection type (1: peak, 2: Root Mean Square).
\end{itemize}

The computation of $g_{x,q_{\theta}}$ is completed as follows.
First, the detection envelope $v$ is obtained from $x, p, \tau_v^{\text{att}}, \tau_v^{\text{rel}}$ as:
\begin{equation}
 v[n] = \sqrt[p]{\beta |x[n]|^p + \bar{\beta} v[n-1]^p}, \quad \text{with } \bar{\beta}=1-\beta
\end{equation}
with $\beta$=$1 - e^{-\frac{2.2}{F_s\tau_v}}$, and 
$\tau_v = \begin{cases}\tau_v^{\text{att}}  & \text{if} \ |x[n]|> v[n-1] \\ \tau_v^{\text{rel}} & \text{otherwise}\end{cases}$
Second, we compute the compression factor $f$ using $v$ and $R$ as:
\begin{equation}
 f[n] = \begin{cases}\left(\frac{l}{v[n]}\right)^{1-\frac{1}{R}} & \text{if} \ v[n] > l, \text{with} \ l=10^{L/20},\\1 & \text{otherwise.}\end{cases}
\end{equation}
Finally, the \ac{drc} gain $g_{x,q_{\theta}}$ is obtained by smoothing the compression factor $f$ using $\tau_g^{\text{att}}$ and $\tau_g^{\text{rel}}$ such as:
\begin{equation}
 g[n] = \gamma f[n] + (1-\gamma) g[n-1],
\end{equation}
with $\gamma = 1 - e^{-\frac{2.2}{F_s\tau_g}}$,
and $\tau_g = \begin{cases} \tau_g^{\text{att}} & \text{if } f[n] > g[n-1] \\ \tau_g^{\text{rel}} & \text{otherwise} \end{cases}$.

The compressed signal is thus obtained using Eq.~\eqref{eq:drc}.
This process corresponds to the implementation included in many existing popular audio software, such as SoX\footnote{SoX Sound eXchange \url{https://sox.sourceforge.net}}.

\subsection{Model-Based DRC Inversion}\label{sec:DRCinv}

The overall method and the corresponding algorithm are basically the same as those given in \cite{GorlowDC}, which show the possibility to estimate $x$ from $y$ when $q_{\theta}$ is exactly known. 

In \cite{GorlowDC}, the authors have defined a $\text{CHARFZERO}()$ function that is based on the Newton-Raphson root finding method to estimate a zero-crossing value $v_0$ of the so-called characteristic function, $\xi_p(v[n])=$
\begin{equation}
(\gamma\kappa v[n]^{-S} + \bar{\gamma}g[n-1])^p (v[n]^p - \bar{\beta} v[n-1]^p)-\beta|y[n]|^p.\label{eq:charf}
\end{equation}
with $S= 1-\frac{1}{R}$, $\kappa=l^S$.

However, the original approach can be computationally expensive, and this is why we propose using the Powell hybrid method instead.
Now, we replace the original $\text{CHARFZERO}$ function with a modification of the Powell hybrid method as implemented in MINPACK \cite{root}, referred to as the ``root" function in the following. This approach provides a faster and more efficient root-finding algorithm, according to our comparative discussion in Section~\ref{sec:opt}. 

Eq.~\eqref{eq:charf} can be used to obtain relevant information about the envelope of $v[n]$.
Hence, when $v[n]$ is known, using $\tilde{x}[n]=v[n]^p$, we can deduce:

\begin{equation}
|x[n]| = \sqrt[p]{\tilde{x}[n] - \frac{1-\beta}{\beta} \tilde{x}[n-1]},
\end{equation}
which enables the estimation of the instantaneous gain required to invert \ac{drc}: $g[n]$=$\frac{|y[n]|}{|x[n]|}$.

\subsection{Proposed Hybrid DRC Inversion Method}
We propose two distinct supervised approaches for DRC parameter estimation.
The classification approach is used when all the possible compressor choices are known. In this case, the DRC parameters $q$ are deduced from a predicted profile label $\theta$, as parameters can be directly deduced from $\theta$. The prediction of the \ac{drc} profile $\theta$ is an audio classification task, since signals compressed by the same profile belong to the same category.
Conversely, when only the compressed signal $y$ is available without prior knowledge, regression becomes necessary for parameter estimation. This method can achieve the same goal, but typically yields lower accuracy than classification.

\subsubsection{\textbf{DRC profile prediction (classification task)}}

\begin{figure}[!t]
    \centering
    \includegraphics[width=1.0\linewidth]{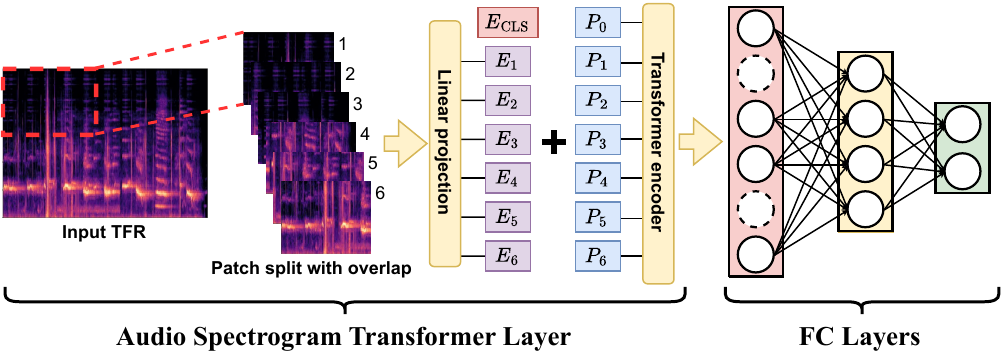}
    \caption{The modified \ac{ast} model architecture, where we add an MLP after the output of the original \ac{ast} model \cite{AST}.}
    \label{fig:ast}
\end{figure}

The architecture of the \ac{ast} \cite{AST} model is illustrated in Figure~\ref{fig:ast}.
First, the input audio signal $y$ is transformed to a time-frequency representation ($\text{TFR}_y$), a spectrogram or mel-spectrogram of $y$, computed using the short-time Fourier transform (STFT). The detailed selection process is described in Section~\ref{sec:results}.
Subsequently, $\text{TFR}_y$ serves as input of the \ac{ast} model.
The representation is split into a sequence of patches with an overlap in both time and frequency dimensions.
Each patch is flattened to a 1D patch embedding $E_i$ using a linear projection layer.
Then, a trainable positional embedding $P_i$ is added to each patch embedding, allowing the model to capture the spatial structure of the 2D audio time-frequency representation.
Additionally, a token \cite{cls} $E_\text{CLS}$ is added at the beginning of the sequence as a classification label. 
We modify the original model by incorporating an MLP at the output. Each hidden layer is followed by a batch normalization and a PReLU activation, enabling prediction of the DRC profile.
We chose the \ac{ast} model because its self-attention mechanism can capture global and local features of the audio spectrogram. ImageNet pre-training provides transferable low-level features. The fine-tuning with additional fully connected layers enhances classification capabilities, and the compressor leaves unique patterns in the spectrogram that are highly consistent with the design goals of the \ac{ast} model.

\subsubsection{\textbf{DRC parameters estimation (regression task)}}

\begin{figure}[!t]
    \centering
    \includegraphics[width=\linewidth]{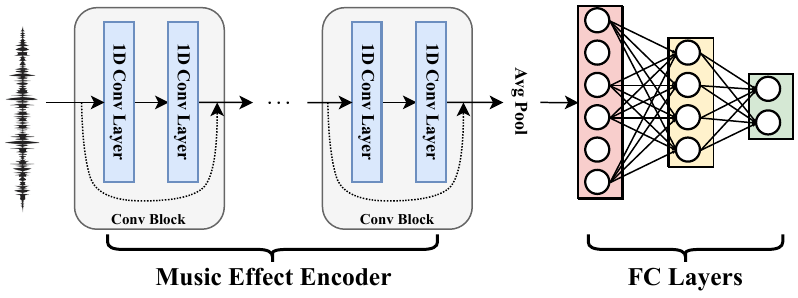}
    \caption{Music Effect Encoder model architecture \cite{beafx}, combined with fully connected layers at the output.}
    \label{fig:mee}
\end{figure}

The \ac{mee} architecture (cf. Figure~\ref{fig:mee}) consists of multiple 1D convolutional blocks, where each block includes two convolutional layers with a residual connection in between. Each convolutional layer is followed by a batch normalization and a rectified linear unit (ReLU) activation function. The output of the last convolutional layer is time-wise encoded through global average pooling, resulting in a dimensionality of 2048, and is used as the music effects feature. The music effect feature is mapped to 512-dimensional features using a linear layer for the contrastive objective, where the loss function used is normalized temperature-scaled cross-entropy loss.
At the output, we use an MLP with 4 layers, respectively of size 2048, 1024, 512, and C, where C is the total number of DRC parameters. Each hidden layer is followed by a batch normalization 1D and a PReLU activation. The output layer, which estimates normalized parameters, is followed by a sigmoid.

\subsubsection{\textbf{End-to-end Architecture}}

\begin{figure*}[!ht]
    \centering
    \includegraphics[width=1.0\textwidth]{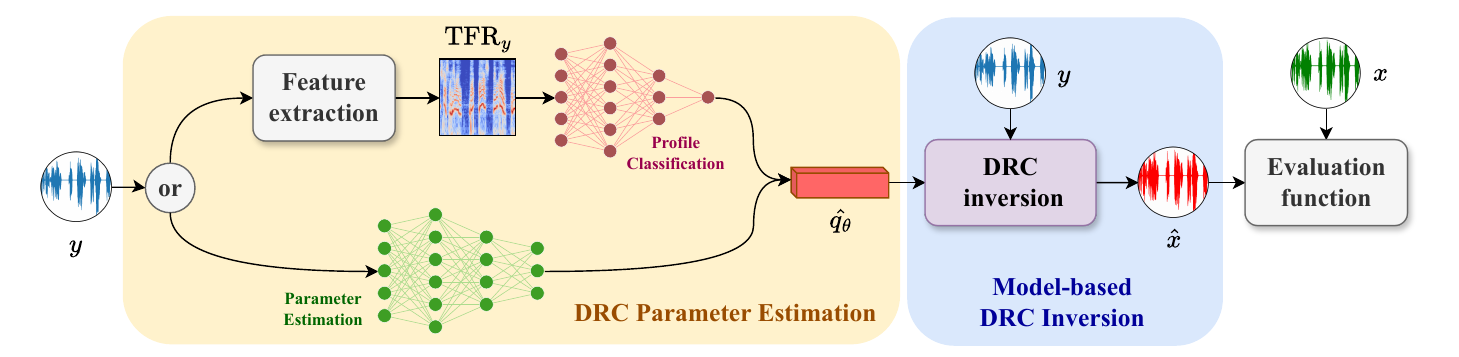}
    \caption{Proposed end-to-end \ac{drc} inversion approach.}
    \label{fig:chain}
\end{figure*}

We propose a new end-to-end \ac{drc} inversion method illustrated in Figure~\ref{fig:chain}. It involves a two-phase approach, which combines a supervised deep neural network architecture with the model-based inversion method presented in Section~\ref{sec:method}. 
Taking the compressed signal $y$ as input, we estimate the DRC parameters in two possible ways:
\begin{itemize}
\item Using the AST model (classification task): This model predicts the profile $\hat{q}_\theta$, for which the set of parameters $q_\theta$ can be deduced (predefined profile),
\item Using the MEE model (regression task): This model directly estimates the parameters $\hat{q}_\theta$ from $y$.
\end{itemize}
Then, the model-based DRC inversion method presented in Section~\ref{sec:DRC} uses the compressed signals $y$ and the resulting parameters $\hat{q}_\theta$ to estimate the original signal $\hat{x}$.

\section{Experiments}\label{sec:exp}
\subsection{Datasets}

We source audio from four different datasets, detailed in Table~\ref{tab:dataset}.
The MedleyDB \cite{MedleyDB} dataset provides the raw tracks of various instruments and vocals. 
We synthesize linear instantaneous mixtures from unprocessed stems and then compress them for training. 
The MUSDB18-HQ \cite{musdb18-hq} and DAFX \cite{DAFX} datasets provide the mixture file for each song, which is the mix of all the music tracks and can be directly used for the evaluation.
Among them, the MUSDB18-HQ dataset provides the original raw signal data, while the DAFX dataset provides mastered signals.
The LibriSpeech \cite{librispeech} dataset provides several versions, among which we choose the one comprising 100 hours of clean speech (train-clean-100).

\begin{table}[!t]
\centering
\renewcommand\arraystretch{1.2}
\caption{Description of the 4 used datasets.}
\resizebox{1.0\linewidth}{!}{
\begin{tabular}{cccc}
\toprule
Name & Data & Sampling Frequency (Hz) & Ref \\ \hline
MedleyDB & Raw & 44100 & \cite{MedleyDB} \\
MUSDB18-HQ & Raw & 44100 & \cite{musdb18-hq} \\
DAFX & Mastered & 32000 & \cite{DAFX} \\
LibriSpeech & Raw & 16000 & \cite{librispeech} \\ 
\bottomrule
\end{tabular}}
\label{tab:dataset}
\end{table}

First, we randomly select $30$ songs from the MedleyDB dataset with a total duration of about $1.6$h. Then, for the other datasets, we select audio with the same total duration.

For all datasets, we randomly select songs with a total duration of about $1.6$h, and then segment all the mixtures into 5-second-length chunks.
We remove the chunk for which the overall root mean square is below -30 dB.
The result is our ground truth original signal $x$.

\begin{table}[!t]
\centering
\renewcommand\arraystretch{1.2}
\caption{The 5 selected DRC profiles for generating the small datasets.}
\resizebox{\linewidth}{!}{
\begin{tabular}{lcccccc}
\toprule
Parameter & Description & A & B & C & D & E \\ \hline
L (dBFS) & Threshold & -32 & -19.9 & -24.4 & -26.3 & -38.0 \\
R ($\text{dB}_{\text{in}}$:$\text{dB}_{\text{out}}$) & Ratio & $3.0:1$ & $1.8:1$ & $3.2:1$ & $7.3:1$ & $4.9:1$ \\
$\tau_{v}^{\text{att}}$ (ms) & Envelope attack & \multirow{2}*{5.0} & \multirow{2}*{5.0} & \multirow{2}*{5.0} & \multirow{2}*{5.0} & \multirow{2}*{5.0} \\
$\tau_{v}^{\text{rel}}$ (ms) & Envelope release &       &       &       &       &       \\
$\tau_{g}^{\text{att}}$ (ms) & Gain attack & 13.0 & 11.0 & 5.8 & 9.0 & 13.1  \\
$\tau_{g}^{\text{rel}}$ (ms) & Gain release & 435 & 49 & 112 & 705 & 257 \\
p ($1$ or $2$) & Detector type & 2 & 2 & 2 & 2 & 2 \\
\bottomrule
\end{tabular}}
\label{tab:5DRC}
\end{table} 

To simulate \ac{drc}, we compress the ground truth audio $x$ with the considered \ac{drc} profiles.
We use 2 groups of \ac{drc} profiles to generate the compressed signal datasets $y$.
We first consider the five distinct DRC profiles in Table~\ref{tab:5DRC}. 
These DRC profiles are standard compressor presets that were previously investigated in \cite{GorlowDC}.
Each original audio clip is compressed using these five compressors. As a result, we get four small datasets corresponding to the compressed signals $y$, respectively referred to as ``Med-6'', ``Mus-6'', ``Dafx-6'', and ``Libri-6''.
Each dataset is made of a total of $6,942$ clips categorized into six distinct classes based on the applied DRC profiles, as well as the neutral signal, labeled as ``0'', which corresponds to the original mixture where no compression is applied.
The signals compressed by the five DRC profiles are labeled as in Table~\ref{tab:5DRC}.

To ensure the generality of the model, we also create four larger datasets using 30 \ac{drc} profiles, which are generated according to the parameters range proposed in \cite{zolzer2002dafx}, presented in Table~\ref{tab:range}.
All the 30 \ac{drc} profiles use an RMS detector, so that $p=2$ for all compressed signals.
Parameter values were linearly spaced and then randomly shuffled to ensure comprehensive coverage of the parameter space while maintaining realistic compression settings.
This set of configurations significantly expands the coverage of parameters such as thresholds, ratios, and attack/release times, providing sufficient complexity to avoid overfitting to a few common configurations.
As a result, we get four larger datasets, each containing $35,867$ samples (about $49.8$ h) of 31 classes of signal (30 DRC profiles + Neutral signal), respectively referred to as ``Med-31'', ``Mus-31'', ``Dafx-31'' and ``Libri-31''.

\begin{table}[!t]
\centering
\renewcommand\arraystretch{1.2}
\caption{The range of parameters for the 30 DRC profiles used to generate the large datasets.}
\resizebox{\linewidth}{!}{
\begin{tabular}{lccc}
\toprule
Parameter & Description & Lower End & Upper End \\ \hline
L (dBFS) & Threshold & -60 & -20 \\
R ($\text{dB}_{\text{in}}$:$\text{dB}_{\text{out}}$) & Ratio & 2 & 15 \\
$\tau_{v}^{\text{att}}$ (ms) & Envelope attack & \multirow{2}{*}{5} & \multirow{2}{*}{130} \\
$\tau_{v}^{\text{rel}}$ (ms) & Envelope release &  &  \\
$\tau_{g}^{\text{att}}$ (ms) & Gain attack & 10 & 500 \\
$\tau_{g}^{\text{rel}}$ (ms) & Gain release & 25 & 2000 \\
p ($1$ or $2$) & Detector type & 2 & 2 \\ 
\bottomrule
\end{tabular}}
\label{tab:range}
\end{table} 

In the following sections, the term ``small datasets'' refers to the four datasets generated with 5 profiles, and the term ``large datasets'' refers to the dataset generated with 30 profiles.

\subsection{Experimental Settings}
 
For both classification and regression experiments, we employ a standard supervised learning protocol with a $4:1$ train-test split ratio. We ensure there is no overlapping chunks between the training and test sets.
We use an Adam optimizer \cite{kingma2014adam} with an initial learning rate of $10^{-4}$ and a batch size of 12 with a maximum of 150 epochs, decreasing the learning rate per epoch with the exponential way by a factor of $\alpha=0.98$.
Training stops after 50 epochs without improvement. In our evaluation, we use the model weights from each configuration that achieves the lowest validation loss during training. 

For the \ac{drc} profile classification task, we minimize the cross-entropy loss between predicted and ground truth labels. Our evaluation considers standard classification metrics, including accuracy, precision, recall, and F1-score, derived from the resulting confusion matrices for the test set. 
For the \ac{drc} parameters estimation regression task, we minimize the mean squared error between the estimated parameters and the ground-truth ones.

All experiments are run on an Intel Xeon CPU with 32 GB of RAM and an Nvidia RTX4080 super GPU with 16GB of VRAM.
Implementation uses the Python language and the Pytorch framework \footnote{Python Github repository: \url{https://github.com/SunHaoRanCN/DRC_Inversion.git}.}

\subsection{Evaluation}

For evaluation, we compute the following metrics:
\begin{itemize}
\item $\mathcal{L}^{\text{MSE}}_{\hat{x},x}$: the Mean Square Error (MSE) between $\hat{x}$ and $x$, computed as:
\begin{equation}
    \mathcal{L}^{\text{MSE}}_{\hat{x},x} = \frac{1}{N}\sum^{N-1}_{n=0}(\hat{x}[n] - x[n])^2
    \label{eq:mse}
\end{equation}
\item $\mathcal{L}^{\text{Mel}}_{\hat{x},x}$: the l2-norm between the log-magnitude mel-spectrogram of $\hat{x}$ and $x$ as in \cite{beafx}:
\begin{equation}
    \mathcal{L}^{\text{Mel}}_{\hat{x}, x} = \|\ln(|\text{Mel}_{\hat{x}}|) - \ln(|\text{Mel}_x|)\|_2
    \label{eq:mel}
\end{equation}
\item SI-SDR: scale-invariant signal-to-distortion-ratio \cite{le2019sdr}, computed as:
\begin{equation}
    \text{SI-SDR}(\hat{x}, x) = 10\log_{10}\left(\frac{\|s_{\text{target}}\|_2^2}{\|\hat{x} - s_{\text{target}}\|_2^2}\right)
\end{equation}
where $s_{\text{target}} = \alpha x$ is the scaled target, and $\alpha = \frac{\hat{x}^Tx}{x^Tx}$ is a scaling factor,
\end{itemize}
Both the original signal $x$ and the estimated signal $\hat{x}$ are normalized by their respective RMS values.

\section{Optimization and Robustness Analysis}\label{sec:opt}

In this section, we evaluate the performance and robustness of the \ac{drc} inversion model.
We first compare the root-finding methods in the CHARFZERO function, showing the advantages of our proposed approach.
We then conduct parameter sensitivity analysis by perturbing each DRC parameter to measure the impact on the reconstruction quality.
Our results highlight the importance of accurate threshold estimation, moderate sensitivity to the ratio parameter, and relative robustness to temporal parameter variations.

\subsection{Function Optimization}
We assess the computational efficiency of the $\text{CHARFZERO}$ function in the \ac{drc} inversion model.
To compare root-finding methods, we conduct a benchmark test using 600 randomly selected 5-second chunks from the MedleyDB's ground truth. These are compressed using 30 DRC profiles Table~\ref{tab:range}, then decompressed using both root-finding variants. We measure the computation time and reconstruction error as Eq.~\eqref{eq:mse}.

Our experiments show that the Newton-Raphson-based $\text{CHARFZERO}$ requires $69.4s$ mean decompression time while the Levenberg-Marquardt method requires $26.1s$. 
This approach also improves the reconstruction quality, with an MSE decreasing from $3.2\times 10^{-5}$ to $8.2\times 10^{-6}$ (Levenberg-Marquardt).

The Levenberg-Marquardt (LM) algorithm implemented in the root function of the scipy library is particularly effective for solving nonlinear least-squares problems. 
It dynamically interpolates between the Gauss-Newton method and gradient-descent-like behavior to ensure both stability and fast convergence.
It minimizes the squared residual norm $\| \xi_p(v) \|^2$, adjusting parameters iteratively to reduce the local error.
By adapting step sizes and directions based on the curvature of $\xi_p(v)$, LM converges efficiently and can achieve high accuracy, especially when close to a solution.



\subsection{Robustness of the DRC Inversion Model}

To investigate the robustness of our \ac{drc} inversion model, we perform parameter sensitivity analysis through controlled perturbations.
Here, we reuse the 600 chunks from the MedleyDB's ground truth, but for efficiency, cut their duration to 1 second.
Using the five distinct DRC profiles from Table~\ref{tab:5DRC}, we compress the signals and apply changes on each parameter within $\pm50\%$ of its reference value while fixing others. For each parameter, we test ten equally spaced variations, comprehensively exploring the parameter space to empirically measure the impact of individual parameter errors on reconstruction quality.

\begin{figure}[!t]
  \centering
  \subfloat[$\mathcal{L}^{\text{MSE}}_{\hat{x},x}$.]{\includegraphics[width=1.0\linewidth]{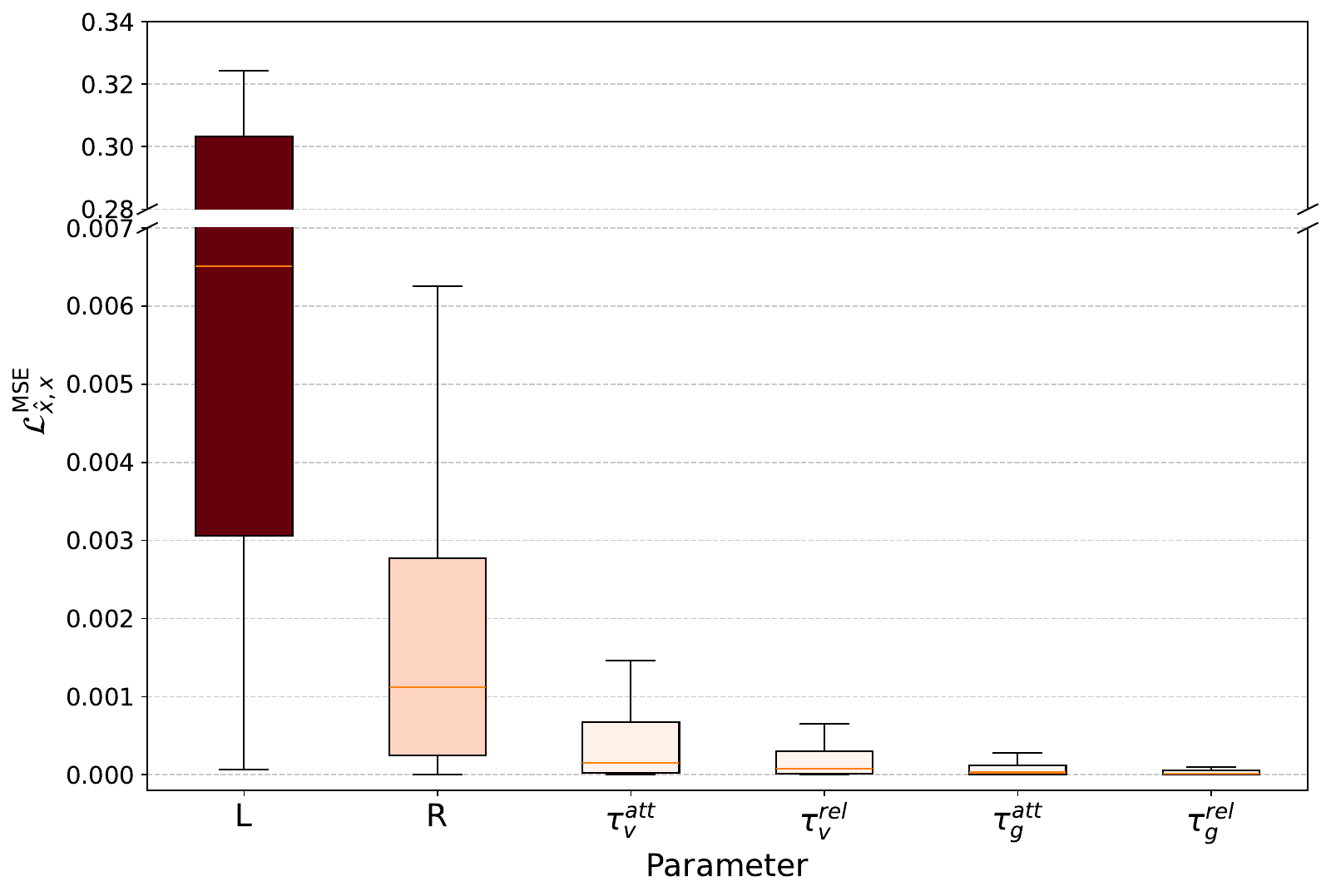} \label{fig:mse_box}}\\
  \subfloat[$\mathcal{L}^{\text{Mel}}_{\hat{x},x}$.]{\includegraphics[width=1.0\linewidth]{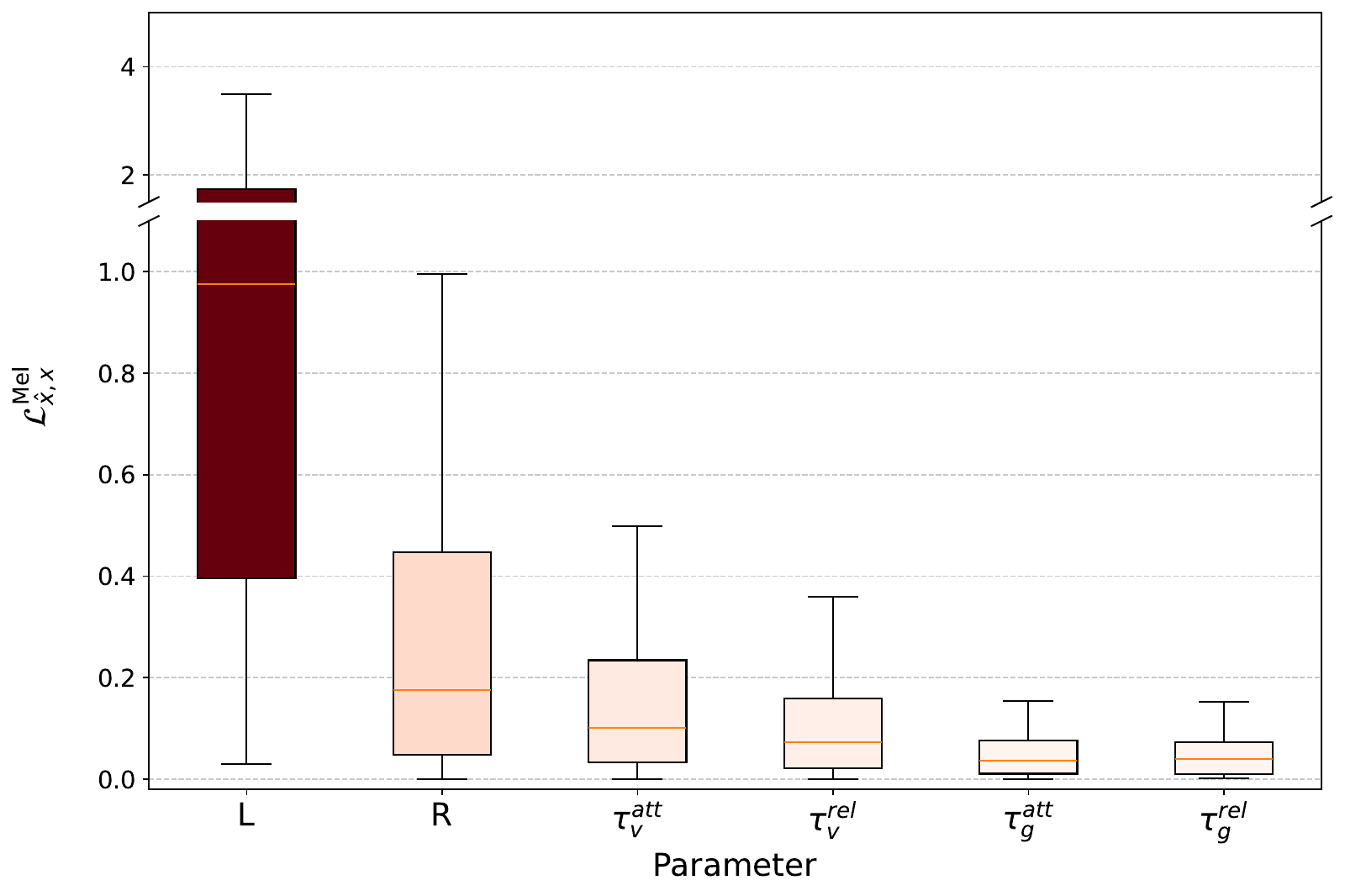} \label{fig:mel_box}}
  \caption{Robustness investigation of \ac{drc} inversion model to parameter variations. Distributions show reconstruction errors when parameters vary within $\pm50\%$ of the ground-truth values.}
  \label{fig:parametr_importance}
\end{figure}

Reconstruction quality is evaluated using Eq.\eqref{eq:mse} and Eq.\eqref{eq:mel}, computed between signals reconstructed with perturbed parameters and the reference (ground truth parameters). Figure~\ref{fig:parametr_importance} visualizes error distributions across all profiles and variations via box plots.

The MSE analysis (Figure~\ref{fig:mse_box}) reveals that the threshold parameter (L) exhibits the highest sensitivity to errors, with median errors around 0.3 and significant variance as indicated by the box plot's whiskers. The ratio parameter (R) shows moderate sensitivity with median errors around 0.005, while the timing parameters (both velocity-based and gain-based) have a remarkably low sensitivity with median errors below 0.002. This hierarchy of parameter sensitivity suggests that accurate threshold estimation is crucial for successful DRC inversion.

The Mel-spectrogram error analysis (Figure~\ref{fig:mel_box}) confirms these observations. The threshold (L) again shows the highest sensitivity, with median errors around 1.0 and extreme values reaching up to 3.5. The ratio (R) maintains its position as the second most sensitive parameter, though with notably lower error magnitudes (median around 0.2). The timing parameters consistently show low sensitivity, with median errors below 0.1.

These observations inform about the design of a robust DRC inversion method, advocating prioritized computational allocation to threshold estimation while permitting greater tolerance for temporal parameter inaccuracies.

\section{Numerical Results}\label{sec:results}

Our experimental results are primarily divided into three parts: First, we fine-tune all the method hyperparameters and settings to maximize the performance of both the \ac{ast} and \ac{mee} models. Second, we present the \ac{drc} inversion results using the complete model. Finally, we compare the performance of our proposed model with several state-of-the-art models.

\subsection{DRC Profile Classification Results}

To the best of our knowledge, there is no study that proposes to use of \ac{ast} for compressed audio signal classification. In this section, we first evaluate its performance for the \ac{drc} profile classification task. Thus, we conduct three comparative experiments to determine the optimal choice of input features, input $\text{TFR}_y$ size, and neural network architecture.

\begin{figure*}[!t]
  \centering
  \subfloat[Classification accuracies of 6 DRC profiles]{\includegraphics[width=0.49\linewidth]{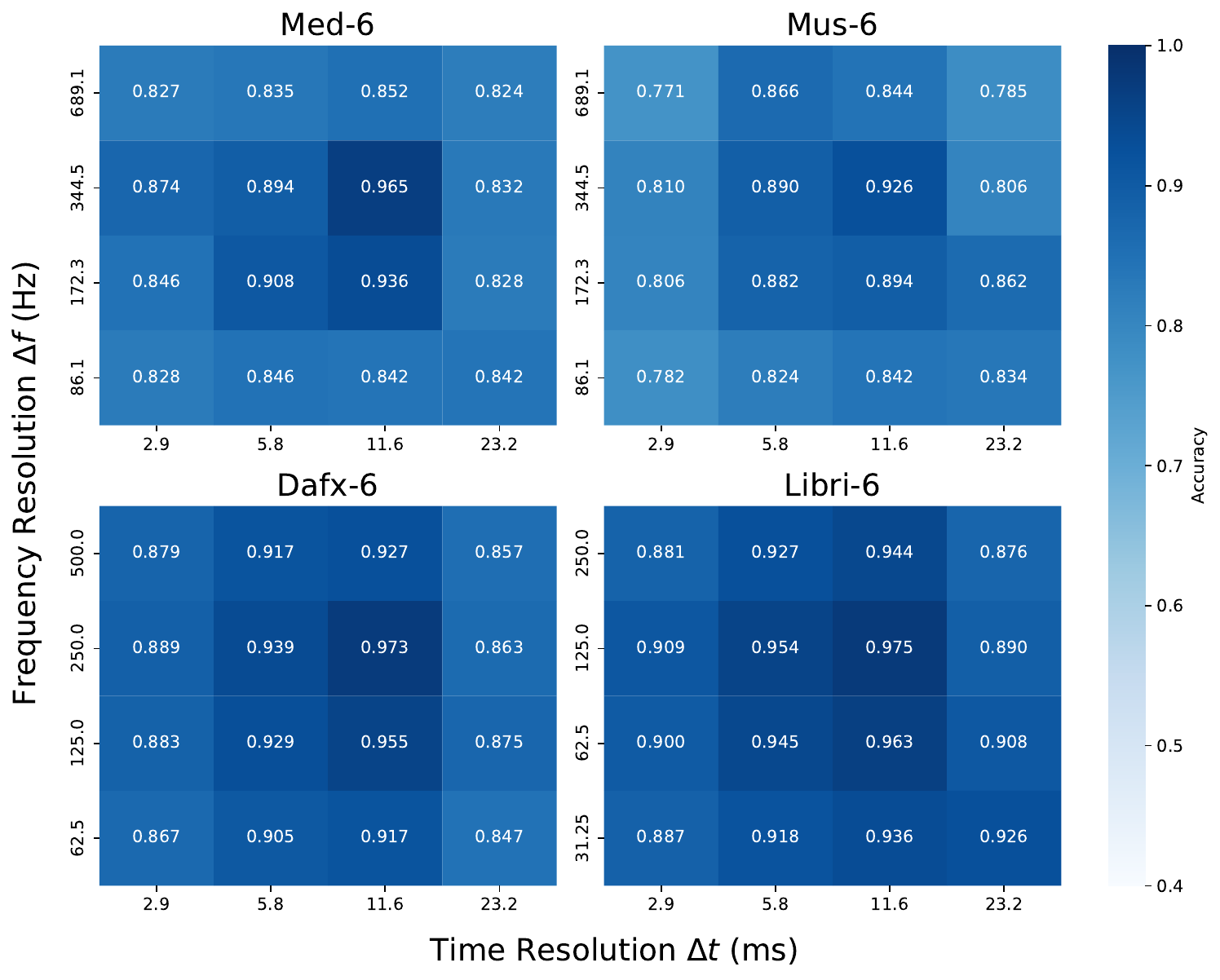}}
  \subfloat[Classification accuracies of 31 DRC profiles]{\includegraphics[width=0.49\linewidth]{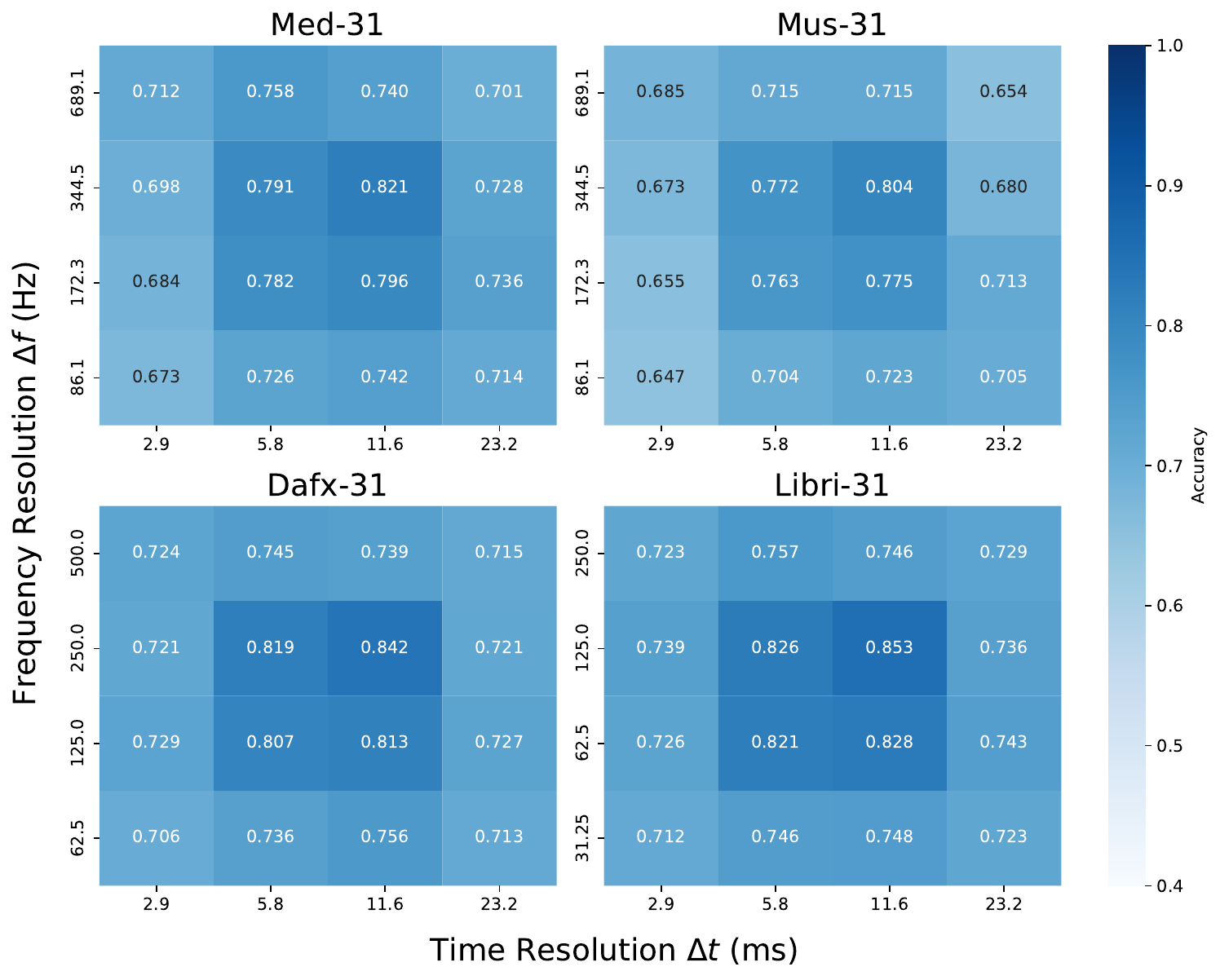}}
  \caption{Classification accuracy heatmaps illustrating the impact of STFT parameters on DRC profile classification using the \ac{ast} model. Each scenario tests four different datasets (Med, Mus, Dafx, and Libri). The x-axis represents the frequency scale resolution in Hz, and the y-axis shows the time frame length in seconds used in the STFT computation.}
  \label{fig:mat_stft}
\end{figure*}

\begin{table*}[!t]
\centering
\caption{Configurations that achieve the highest accuracy on the test set using different input $\text{TFR}_y$ with the original \ac{ast} model.}
\renewcommand\arraystretch{1.2}
\begin{subtable}[!t]{.49\textwidth}
\centering
\caption{Configuration exploration results for small datasets}
\vspace{0pt}
\resizebox{\textwidth}{!}{
\begin{tabular}{ccccc}
\toprule
Dataset & Input & Size & Time/epoch(h) & Acc \\ \hline
\multirow{4}{*}{Med-6} & MFCC & $64 \times 431$ & 0.030 & 0.79 \\
 & STFT & $64 \times 431$ & \textbf{0.008} & \textbf{0.96} \\
 & MelS & $128 \times 431$ & 0.021 & 0.95 \\
 & CQT & $64 \times 862$ & 0.26 & 0.73 \\ \hline
\multirow{4}{*}{Mus-6} & MFCC & $64 \times 431$ & 0.030 & 0.76 \\
 & STFT & $64 \times 431$ & \textbf{0.008} & \textbf{0.92} \\
 & MelS & $128 \times 431$ & 0.021 & 0.91 \\
 & CQT & $64 \times 862$ & 0.26 & 0.71 \\ \hline
\multirow{4}{*}{Dafx-6} & MFCC & $64 \times 431$ & 0.028 & 0.88 \\
 & STFT & $64 \times 431$ & \textbf{0.007} & \textbf{0.97} \\
 & MelS & $64 \times 431$ & 0.019 & \textbf{0.97} \\
 & CQT & $128 \times 862$ & 0.23 & 0.82 \\ \hline
\multirow{4}{*}{Libri-6} & MFCC & $64 \times 431$ & 0.024 & 0.90 \\
 & STFT & $64 \times 431$ & \textbf{0.006} & \textbf{0.98} \\
 & MelS & $64 \times 431$ & 0.017 & \textbf{0.98} \\
 & CQT & $128 \times 862$ & 0.20 & 0.83 \\
\bottomrule 
\end{tabular}}
\end{subtable}%
\begin{subtable}[!t]{.49\textwidth}
\centering
\caption{Configuration exploration results for large datasets}
\vspace{0pt}
\resizebox{\textwidth}{!}{
\begin{tabular}{ccccc}
\toprule
Dataset & Input & Size & Time/epoch(h) & Acc \\[0.02em] \hline
\multirow{4}{*}{Med-31} & MFCC & $128 \times 431$ & 0.093 & 0.66 \\[0.035em]
 & STFT & $64 \times 431$ & \textbf{0.037} & \textbf{0.82} \\[0.035em]
 & MelS & $128 \times 431$ & 0.074 & 0.81 \\[0.02em]
 & CQT & $128 \times 862$ & 0.79 & 0.55 \\ \hline
\multirow{4}{*}{Mus-31} & MFCC & $128 \times 431$ & 0.093 & 0.65 \\[0.035em]
 & STFT & $64 \times 431$ & \textbf{0.037} & \textbf{0.80} \\[0.035em]
 & MelS & $128 \times 431$ & 0.074 & 0.78 \\[0.02em]
 & CQT & $128 \times 862$ & 0.79 & 0.53 \\ \hline
\multirow{4}{*}{Dafx-31} & MFCC & $64 \times 431$ & 0.091 & 0.72 \\[0.03em]
 & STFT & $64 \times 431$ & \textbf{0.036} & \textbf{0.84} \\[0.03em]
 & MelS & $64 \times 431$ & 0.072 & 0.83 \\[0.02em]
 & CQT & $64 \times 862$ & 0.75 & 0.66 \\ \hline
\multirow{4}{*}{Libri-31} & MFCC & $64 \times 431$ & 0.089 & 0.73 \\[0.03em]
 & STFT & $64 \times 431$ & \textbf{0.035} & \textbf{0.85} \\[0.03em]
 & MelS & $64 \times 431$ & 0.070 & 0.84 \\[0.02em]
 & CQT & $64 \times 862$ & 0.71 & 0.67 \\
\bottomrule 
\end{tabular}}
\end{subtable}
\label{tab:tf_compare}
\end{table*}

\subsubsection{\textbf{Input features and size selection}}

Inspired by \cite{huzaifah2017comparison}, we empirically compare four commonly used time-frequency representations ($\text{TFR}_y$): \ac{mfcc}, Short-Time Fourier Transform (STFT) spectrogram, Mel Spectrogram (MelS), and Constant-Q Transform (CQT) spectrogram. 
We evaluate their impact on the classification performance of the \ac{drc} profiles. All time-frequency features are computed using \textit{librosa} \cite{mcfee2015librosa}. Although \ac{drc} is a time-domain transform, it induces changes in both the time and frequency domains. 
Therefore, it is essential to investigate the impact of different $\text{TFR}_y$ sizes (time and frequency resolutions) on the experimental outcomes.

To ensure a comprehensive evaluation, we explore various input feature sizes to identify the optimal choice. 
Through our experiments, we observe that variations in the time domain have a more significant impact than those in the frequency domain. 
Therefore, we select time-frequency feature sizes with larger time domain dimensions and smaller frequency domain dimensions for comparison

Figure~\ref{fig:mat_stft} illustrates some of the input sizes yielding the best performance. For the STFT, a discrete Fourier transform with a sliding Hann window is applied to overlapping signal segments. For the frequency scale, we explore four different bin sizes, respectively 32, 64, 128, and 256, corresponding to the x-axis in the subgraph from right to left. For the time scale, we also examine four different frame counts, respectively 216, 431, 862, and 1,723, corresponding to the y-axis in the subgraph from top to bottom. We investigate their combinations, assessing the impact of 16 distinct input feature sizes on classification accuracy. 
Table~\ref{tab:tf_compare} summarizes the experimental configurations and results achieving the highest classification accuracy.

From Figure~\ref{fig:mat_stft}, in the small datasets experiments, all four tested datasets achieve notably high classification accuracy on the test set, with peak values exceeding $0.95$. The Libri-6 and Dafx-6 datasets show particularly robust performance, consistently maintaining accuracy above $0.90$ across most parameter combinations. Optimal performance is observed with intermediate time frame lengths ($5.8$ ms to $11.6$ ms) and frequency bins (64 and 128), where classification accuracy reached 0.975 for the Libri-6 dataset. This suggests that moderate temporal and frequency resolutions provide sufficient discriminative features for distinguishing between different DRC profiles in the 6-\ac{drc}-profiles scenario.

The large datasets' results reveal a more challenging classification task, with overall lower accuracy values ranging from $0.65$ to $0.85$. This decrease in performance is due to the increased complexity of distinguishing among 31 different DRC profiles. The optimal parameter combinations remained consistent with those of the small datasets. Notably, the Libri-31 and Dafx-31 datasets maintained relatively better performance compared to Med-31 and Mus-31, suggesting that these audio sources may contain more distinctive characteristics relevant for DRC profile classification.

A common pattern observed across both dataset types is the degradation in performance at very short or very long time frame lengths. This indicates that extremely fine or coarse temporal resolutions may either capture excessive noise or overlook important dynamic characteristics of the DRC processing. Additionally, increasing the number of frequency bins beyond 128 generally did not yield significant improvements in classification accuracy, suggesting that a moderate frequency resolution is sufficient for this task.

Through our exploration of the DRC profile classification task, for both small and large datasets, using the STFT as the input feature yields the highest classification accuracy, with the input feature size of $64 \times 431$ \footnote{The size $a \times b$ corresponds to the number of frequency bins and time instants, respectively.} achieving the best results. The classification accuracy when using MelS as the input feature is comparable to that of STFT, particularly for several small datasets. However, the overall performance of MelS is slightly lower than that of STFT, and the required training time is significantly longer, so it is not used in this work. Furthermore, the classification accuracies when using MFCC or CQT as input features are lower, and the required training times are longer, so these features are not considered in this work.

\subsubsection{\textbf{Choice of the number of FC layers:}}

\begin{figure*}[!t]
  \centering
  \subfloat[Results for small datasets.]{\includegraphics[width=0.49\linewidth]{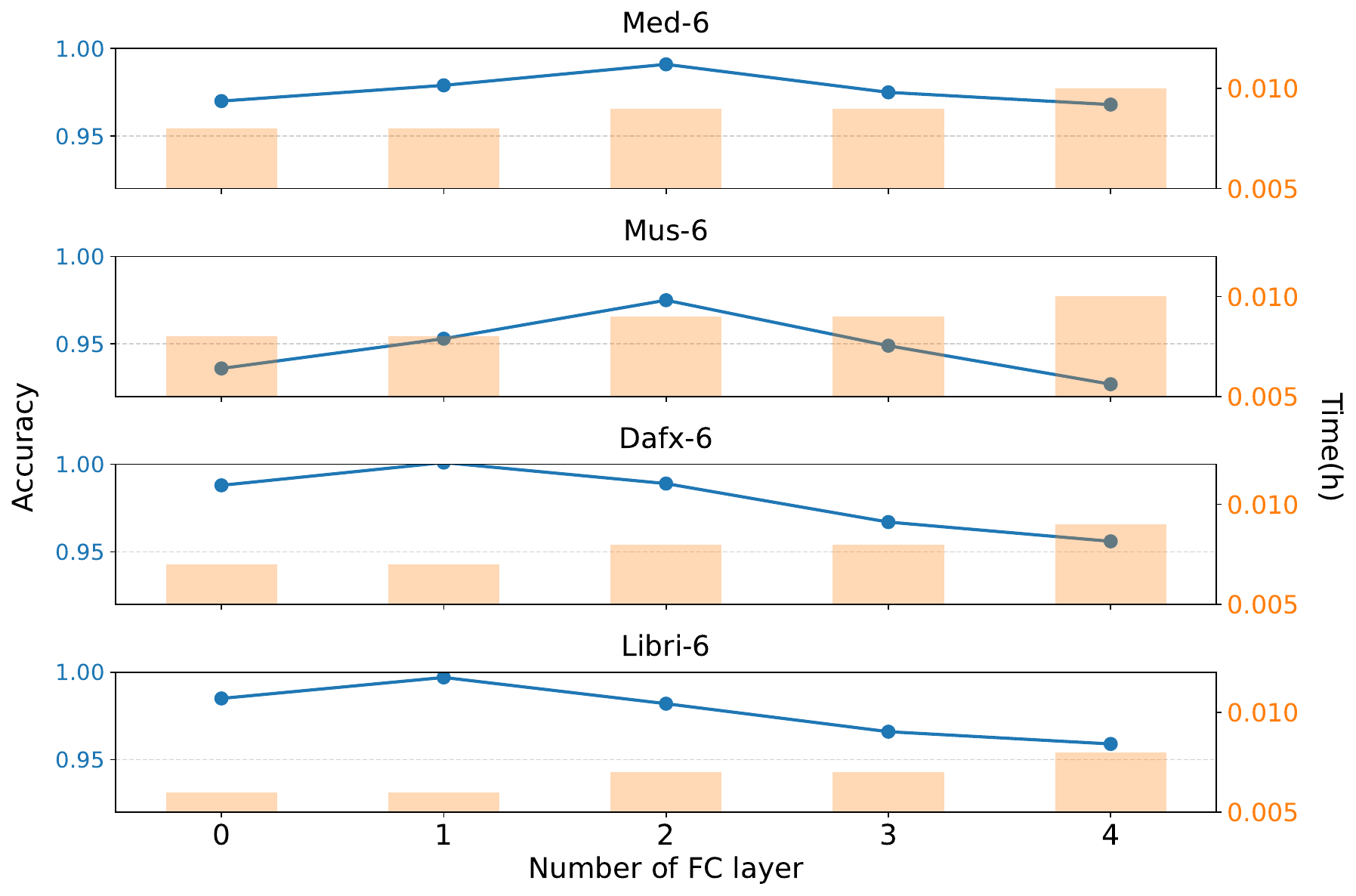} \label{fig:layers_small}}
  \subfloat[Results for large datasets.]{\includegraphics[width=0.49\linewidth]{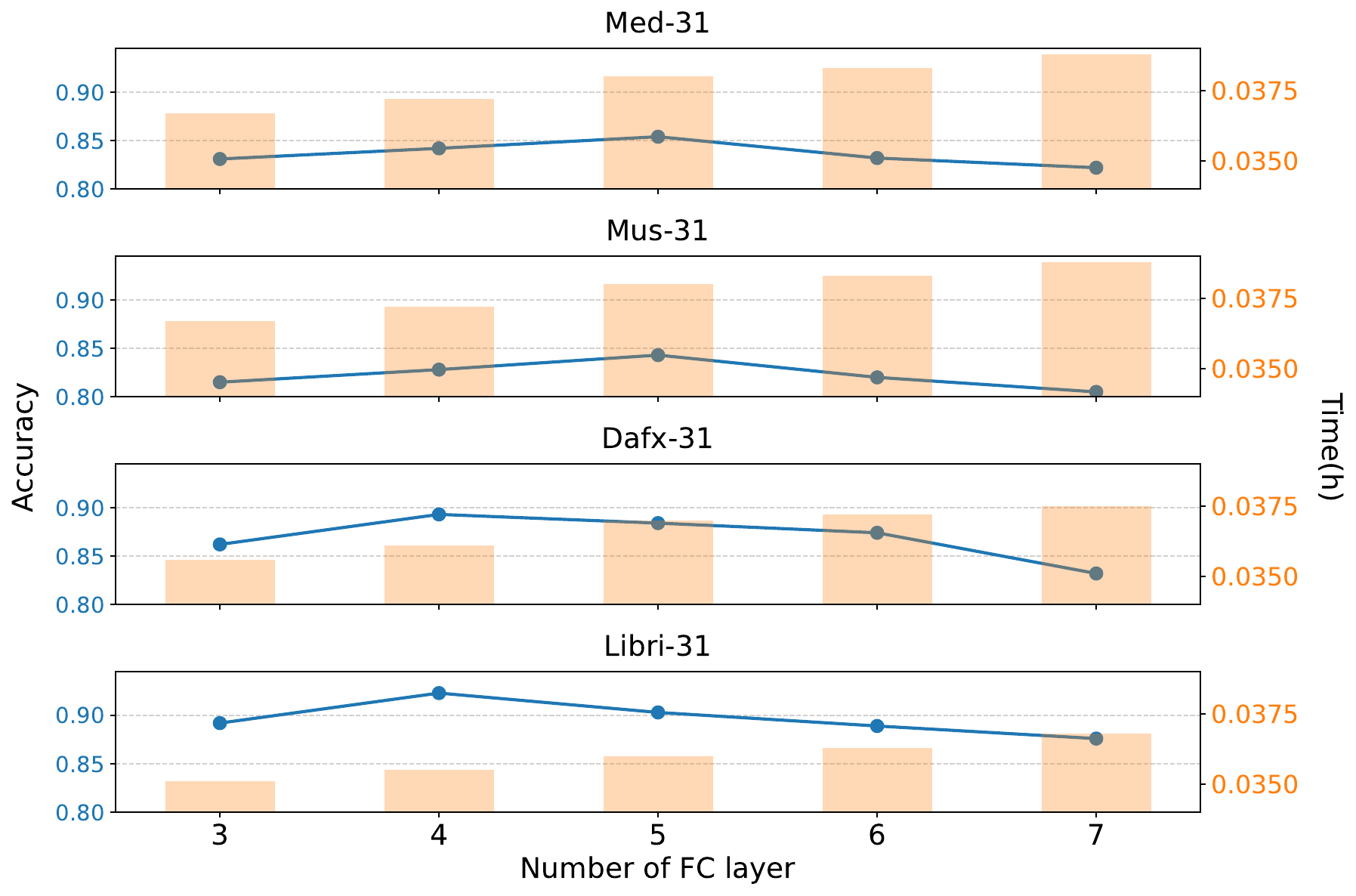} \label{fig:layers_large}}
  \caption{Impact of the number of FC layers on DRC profile classification using the \ac{ast} model. The blue lines indicate classification accuracy (left y-axis), while orange bars show training time per epoch in hours (right y-axis). }
  \label{fig:layers}
\end{figure*}


As mentioned above, we add an MLP after the output of the original \ac{ast} model to enhance classification accuracy.
Varying the number of Fully Connected (FC) layers in the MLP allows us to assess its ability to learn intricate patterns within the STFT features. 
Consequently, testing different numbers of FC layers in the MLP is essential to find the best architecture.

Specifically, for small datasets, given the lower complexity of the task, we select FC layers configurations, respectively of $0$, $1$, $2$, $3$, and $4$, where $0$ layer corresponds to the original \ac{ast} model without additional FC layer.
For large datasets, given the increased number of \ac{drc} profiles and higher task complexity, we explore FC layer configurations, respectively of $3$, $4$, $5$, $6$, and $7$. Based on the experimental results referenced earlier, the input feature size for both large and small datasets is consistently set to $64 \times 431$.

\begin{table}[!t]
\centering
\caption{Highest test set accuracy and corresponding \ac{ast} model configurations for each dataset in the \ac{drc} profile classification task, with input size set to $64 \times 431$.}
\renewcommand\arraystretch{1.2}
\begin{subtable}[t]{.49\linewidth}
\centering
\caption{Results for small datasets.}
\resizebox{1.0\linewidth}{!}{
\begin{tabular}{cccc}
\toprule 
Dataset & FC-layer & Time/epoch(h) & Acc \\ 
\hline
Med-6 & 2 & 0.04 & 0.986 \\
Mus-6 & 2 & 0.04 & 0.965 \\
Dafx-6 & 1 & 0.03 & 0.996 \\
Libri-6 & 1 & 0.03 & 0.997 \\
\bottomrule 
\end{tabular}}
\end{subtable}
\begin{subtable}[t]{.49\linewidth}
\centering
\caption{Results for large datasets.}
\resizebox{1.0\linewidth}{!}{
\begin{tabular}{ccccc}
\toprule 
Dataset & FC-layer & Time/epoch(h) & Acc \\ 
\hline
Med-31 & 5 & 0.09 & 0.854 \\[0.035em]
Mus-31 & 5 & 0.09 & 0.843 \\[0.035em]
Dafx-31 & 4 & 0.08 & 0.893 \\[0.02em]
Libri-31 & 4 & 0.08 & 0.923 \\
\bottomrule 
\end{tabular}}
\end{subtable}
\label{tab:ast_best_all}
\end{table}

From Figure~\ref{fig:layers_small}, for the Med-6 and Mus-6 datasets, using 2 FC layers in the MLP yields the highest accuracy, whereas for the Dafx-6 and Libri-6 datasets, 1 FC layer is optimal. In simpler tasks, increasing the number of FC layers increases model complexity, which can reduce the classification performance. 
Conversely, from Figure~\ref{fig:layers_large}, for the Med-31 and Mus-31 datasets, 5 FC layers in the MLP provide the highest accuracy, while for the Dafx-31 and Libri-31 datasets, 4 FC layers are the best choice. This suggests that more complex tasks require deeper models to capture fine-grained features, and increasing the number of layers can effectively enhance the classification performance. Across all tasks, training time per epoch rises with the number of FC layers, aligning with expectations.


Table~\ref{tab:ast_best_all} summarizes these experimental results. Generally, for the \ac{drc} profile classification task, we use the STFT as the input feature. For small datasets, one or two FC layers allow to achieve high classification performance, with accuracies ranging from 95\% to 99\%. For large datasets, given the increased task complexity, four or five FC layers are necessary for optimal performance, yielding accuracies between 85\% and 92\%.

\subsection{DRC Parameters Estimation Results}

We propose to use the MEE \cite{beafx} regression model to estimate audio effect parameters from mastered signals.
Note that, since the detector type of all DRC profiles in this work is RMS (i.e., the seventh parameter $p=2$), we only need to estimate the six other parameters.
Therefore, in this work, we directly use the same model configuration as in \cite{beafx} and we assess the performance of the \ac{mee} model on the DRC parameters estimation task.

First, we present in Table~\ref{tab:mee} the experimental results of DRC parameters estimation using the \ac{mee} model. 
Following our exploration of the \ac{ast} model, we initially use small datasets to validate the performance of the \ac{mee} model on the DRC parameter estimation task and subsequently test the model's generalization ability on larger datasets. 
Additionally, we use the Time-Frequency Encoder (TFE) \cite{beafx}, as a comparison model to ensure that the \ac{mee} model is the better choice. 
The input to the TFE model is the STFT of the signal, with the STFT size set to $64 \times 431$. In this task, we minimize $\mathcal{L}^{\text{MSE}}_{\hat{q}_{\theta},q_{\theta}}$ computed using Eq.~(1) for a total number of parameters set to 6.

\begin{table}[!t]
\centering
\caption{Comparison of DRC parameters estimation results between the \ac{mee} and TFE models. The input feature for the TFE model is the STFT of size $64 \times 431$.}
\label{tab:mee}
\renewcommand\arraystretch{1.2}
\begin{subtable}[t]{0.495\linewidth}
\centering
\caption{MEE, small datasets.}
\resizebox{1.0\linewidth}{!}{
\begin{tabular}{ccc}
\toprule 
Dataset & Time/epoch(h) & $\mathcal{L}^{\text{MSE}}_{\hat{q_{\theta}},q_{\theta}}$ \\ \hline
Med-6 & 0.015 & \textbf{0.007} \\
Mus-6 & 0.016 & \textbf{0.010} \\
Dafx-6 & 0.013 & \textbf{0.006} \\
Libri-6 & 0.008 & \textbf{0.003} \\
\bottomrule 
\end{tabular}}
\end{subtable}%
\begin{subtable}[t]{0.495\linewidth}
\centering
\caption{MEE, large datasets.}
\resizebox{1.0\linewidth}{!}{
\begin{tabular}{ccc}
\toprule 
Dataset & Time/epoch(h) & $\mathcal{L}^{\text{MSE}}_{\hat{q_{\theta}},q_{\theta}}$ \\ [0.055em]\hline
Med-31 & 0.084 & \textbf{0.031} \\[0.04em]
Mus-31 & 0.087 & \textbf{0.039} \\[0.04em]
Dafx-31 & 0.065 & \textbf{0.025} \\[0.04em]
Libri-31 & 0.040 & \textbf{0.012} \\[0.035em]
\bottomrule 
\end{tabular}}
\end{subtable}

\vspace{0.3cm}

\begin{subtable}[t]{0.495\linewidth}
\centering
\caption{TFE, small datasets.}
\resizebox{1.0\linewidth}{!}{
\begin{tabular}{ccc}
\toprule 
Dataset & Time/epoch(h) & $\mathcal{L}^{\text{MSE}}_{\hat{q}_{\theta}, q_{\theta}}$ \\ \hline
Med-6 & 0.0054 & 0.055 \\
Mus-6 & 0.0056 & 0.056 \\
Dafx-6 & 0.0042 & 0.048 \\
Libri-6 & 0.0027 & 0.046 \\
\bottomrule 
\end{tabular}}
\end{subtable}%
\begin{subtable}[t]{0.495\linewidth}
\centering
\caption{TFE, large datasets.}
\resizebox{1.0\linewidth}{!}{
\begin{tabular}{ccc}
\toprule 
Dataset & Time/epoch(h) & $\mathcal{L}^{\text{MSE}}_{\hat{q_{\theta}},q_{\theta}}$ \\ [0.055em]\hline
Med-31 & 0.032 & 0.058 \\[0.04em]
Mus-31 & 0.038 & 0.058 \\[0.04em]
Dafx-31 & 0.022 & 0.054 \\[0.04em]
Libri-31 & 0.015 & 0.052 \\[0.035em]
\bottomrule 
\end{tabular}}
\end{subtable}
\label{tab:reg_compare}
\end{table}

Table~\ref{tab:mee} presnets the experimental results.
Since the input signal length is 5 seconds, the difference in training time between small and large datasets depends solely on the number of samples and the sampling frequency.
According to Table~\ref{tab:reg_compare}, the MEE model obtains better results in comparison to the TFE in terms of regression accuracy for all the datasets.
In terms of computational efficiency, the TFE model exhibits faster training times, though the difference between the two is not substantial.
Additionally, for both small and large datasets, the \ac{mee} model achieves the smallest parameter error on the LibriSpeech dataset and the largest on the MUSDB18-HQ dataset.

Combining the results from Table~\ref{tab:ast_best_all} and Table~\ref{tab:mee}, we observe that increasing the number of DRC profiles significantly reduces the accuracy of the classification task and increases the error in the regression task.
Given that the experiments with small datasets are relatively straightforward, the generalization experiments on larger datasets markedly increase complexity.
We consider the observed decline in performance to be reasonable and acceptable.


\subsection{Performance Optimization}

\begin{figure}[!t]
    \centering
    \includegraphics[width=1.0\linewidth]{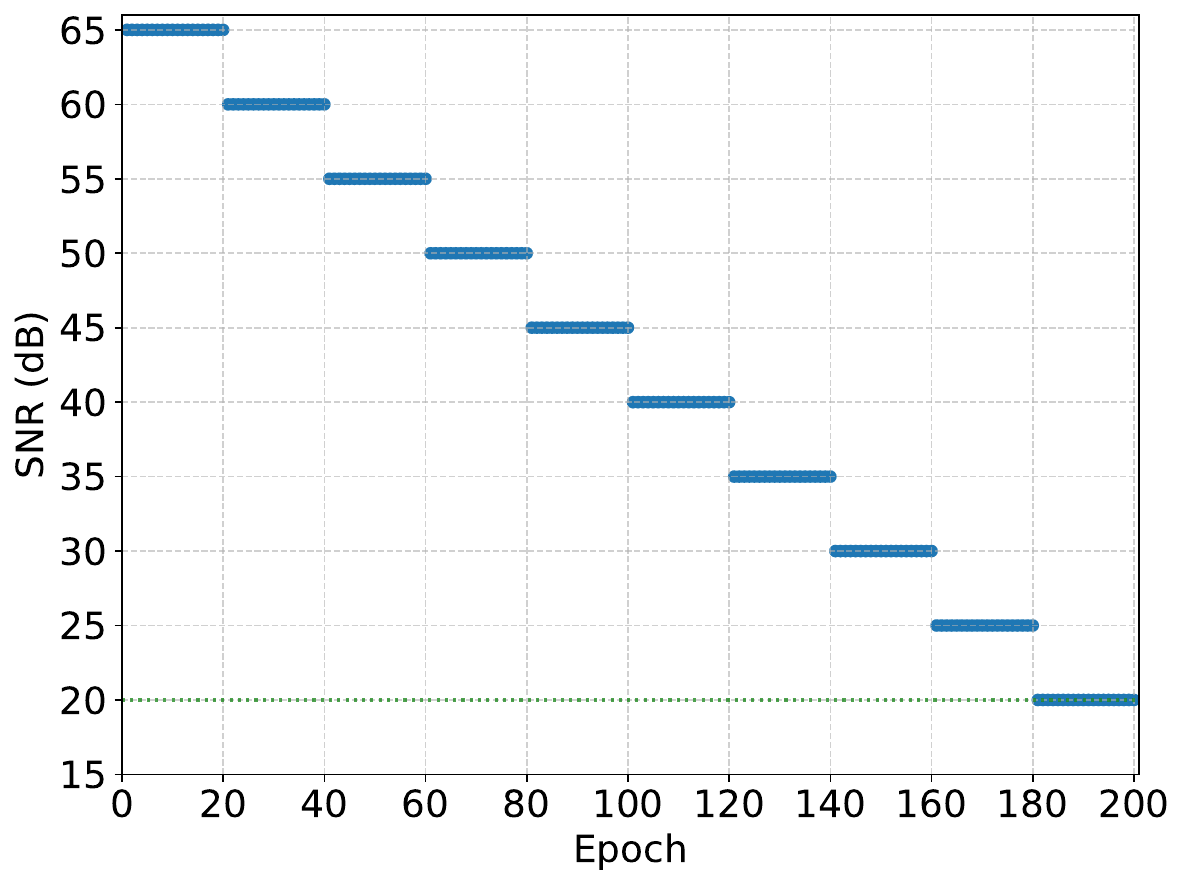}
    \caption{Curriculum Learning Strategy for Training.}
    \label{fig:snr}
\end{figure}

\begin{table}[!t]
\centering
\caption{Comparison of \ac{drc} Profiles Classification Results between with \& without data augmentation.}
\label{tab:ast_aug}
\begin{subtable}[t]{0.495\linewidth}
\centering
\resizebox{1.0\linewidth}{!}{
\begin{tabular}{ccc}
\toprule 
Dataset & Time/epoch(h) & Acc \\ 
\hline
Med-6 & 0.04 & 0.986 \\
Mus-6 & 0.04 & 0.965 \\
Dafx-6 & 0.03 & 0.996 \\
Libri-6 & 0.03 & 0.997 \\
\bottomrule 
\end{tabular}}
\caption{small, without aug}
\end{subtable}%
\begin{subtable}[t]{0.495\linewidth}
\centering
\resizebox{1.0\linewidth}{!}{
\begin{tabular}{ccc}
\toprule 
Dataset & Time/epoch(h) & Acc \\ 
\hline
Med-6 & 0.10 & \textbf{0.992} \\[0.035em]
Mus-6 & 0.10 & \textbf{0.988} \\[0.035em]
Dafx-6 & 0.08 & \textbf{0.997} \\[0.035em]
Libri-6 & 0.07 & \textbf{0.998} \\[0.035em]
\bottomrule 
\end{tabular}}
\caption{small, with aug}
\end{subtable}
\begin{subtable}[t]{0.495\linewidth}
\centering
\resizebox{1.0\linewidth}{!}{
\begin{tabular}{ccc}
\toprule 
Dataset & Time/epoch(h) & Acc \\ 
\hline
Med-31 & 0.09 & 0.854 \\
Mus-31 & 0.09 & 0.843 \\
Dafx-31 & 0.08 & 0.893 \\
Libri-31 & 0.08 & 0.923 \\
\bottomrule 
\end{tabular}}
\caption{large, without aug}
\end{subtable}%
\begin{subtable}[t]{0.495\linewidth}
\centering
\resizebox{1.0\linewidth}{!}{
\begin{tabular}{ccc}
\toprule 
Dataset & Time/epoch(h) & Acc \\ 
\hline
Med-31 & 0.19 & \textbf{0.886} \\[0.035em]
Mus-31 & 0.19 & \textbf{0.872} \\[0.035em]
Dafx-31 & 0.17 & \textbf{0.935} \\[0.035em]
Libri-31 & 0.16 & \textbf{0.957} \\[0.035em]
\bottomrule 
\end{tabular}}
\caption{large, with aug}
\end{subtable}
\end{table}

\begin{table}[!t]
\centering
\caption{Comparison of \ac{drc} Parameter Estimation Results, between with \& without data augmentation.}
\label{tab:mee_aug}
\begin{subtable}[t]{0.495\linewidth}
\centering
\resizebox{1.0\linewidth}{!}{
\begin{tabular}{ccc}
\toprule 
Dataset & Time/epoch(h) & $\mathcal{L}^{\text{MSE}}_{\hat{q_{\theta}},q_{\theta}}$ \\ 
\hline
Med-6 & 0.015 & 0.007 \\
Mus-6 & 0.016 & 0.010 \\
Dafx-6 & 0.013 & 0.006 \\
Libri-6 & 0.008 & 0.003 \\
\bottomrule 
\end{tabular}}
\caption{small, without aug}
\end{subtable}%
\begin{subtable}[t]{0.495\linewidth}
\centering
\resizebox{1.0\linewidth}{!}{
\begin{tabular}{ccc}
\toprule 
Dataset & Time/epoch(h) & $\mathcal{L}^{\text{MSE}}_{\hat{q_{\theta}},q_{\theta}}$ \\ 
\hline
Med-6 & 0.032 & \textbf{0.0032} \\[0.035em]
Mus-6 & 0.032 & \textbf{0.0054} \\[0.035em]
Dafx-6 & 0.027 & \textbf{0.0015} \\[0.035em]
Libri-6 & 0.019 & \textbf{0.0008} \\[0.035em]
\bottomrule 
\end{tabular}}
\caption{small, with aug}
\end{subtable}
\begin{subtable}[t]{0.495\linewidth}
\centering
\resizebox{1.0\linewidth}{!}{
\begin{tabular}{ccc}
\toprule 
Dataset & Time/epoch(h) & $\mathcal{L}^{\text{MSE}}_{\hat{q_{\theta}},q_{\theta}}$ \\ 
\hline
Med-31 & 0.084 & 0.031 \\
Mus-31 & 0.087 & 0.039 \\
Dafx-31 & 0.065 & 0.025 \\
Libri-31 & 0.040 & 0.012 \\
\bottomrule 
\end{tabular}}
\caption{large, without aug}
\end{subtable}%
\begin{subtable}[t]{0.495\linewidth}
\centering
\resizebox{1.0\linewidth}{!}{
\begin{tabular}{ccc}
\toprule 
Dataset & Time/epoch(h) & $\mathcal{L}^{\text{MSE}}_{\hat{q_{\theta}},q_{\theta}}$ \\ 
\hline
Med-31 & 0.17 & \textbf{0.011} \\[0.035em]
Mus-31 & 0.18 & \textbf{0.013} \\[0.035em]
Dafx-31 & 0.14 & \textbf{0.008} \\[0.035em]
Libri-31 & 0.08 & \textbf{0.005} \\[0.035em]
\bottomrule 
\end{tabular}}
\caption{large, with aug}
\end{subtable}
\end{table}

To enhance model performance and generalization, we implement a training strategy integrating data augmentation and curriculum learning \cite{CL}.

For data augmentation, we double the training dataset size by injecting Gaussian noise into the original audio chunks, generating noisy samples $y_{\text{noisy}}[n]$:
\begin{equation}
y_{\text{noisy}}[n] = y[n] + b[n], \quad b[n] \sim \mathcal{N}(0, \sigma^2)
\end{equation}
where $\sigma^2 = \frac{P_{\text{signal}}}{10^{\text{SNR}_{\text{dB}}/10}}$, $P_{\text{signal}} = \frac{1}{N} \sum_{n=1}^{N} |y[n]|^2$. As a result, clean and noisy signals each constitute half of the augmented dataset. 

Concurrently, we use curriculum learning to improve noise-level adaptability by progressively decreasing the signal-to-noise ratio (SNR) during training. Figure~\ref{fig:snr} details the SNR reduction schedule.
Starting from 65 dB, the SNR value decreases by 5 dB every 20 epochs of training until it decreases to 20 dB.

Tables~\ref{tab:ast_aug} and \ref{tab:mee_aug} present experimental results for DRC profile classification and parameter estimation, respectively, comparing performance with and without augmentation. All experiments use optimal configurations identified previously.

Results indicate that augmentation enhances performance across all conditions, particularly for challenging large-dataset tasks. Both tasks exhibit consistent computational overhead from augmentation, with training time approximately doubling. However, significant performance gains, especially in complex configurations, justify this cost for high-accuracy applications.

\subsection{DRC Parameters Visualization}

\begin{figure*}[!t]
    \centering
    \includegraphics[width=1.0\linewidth]{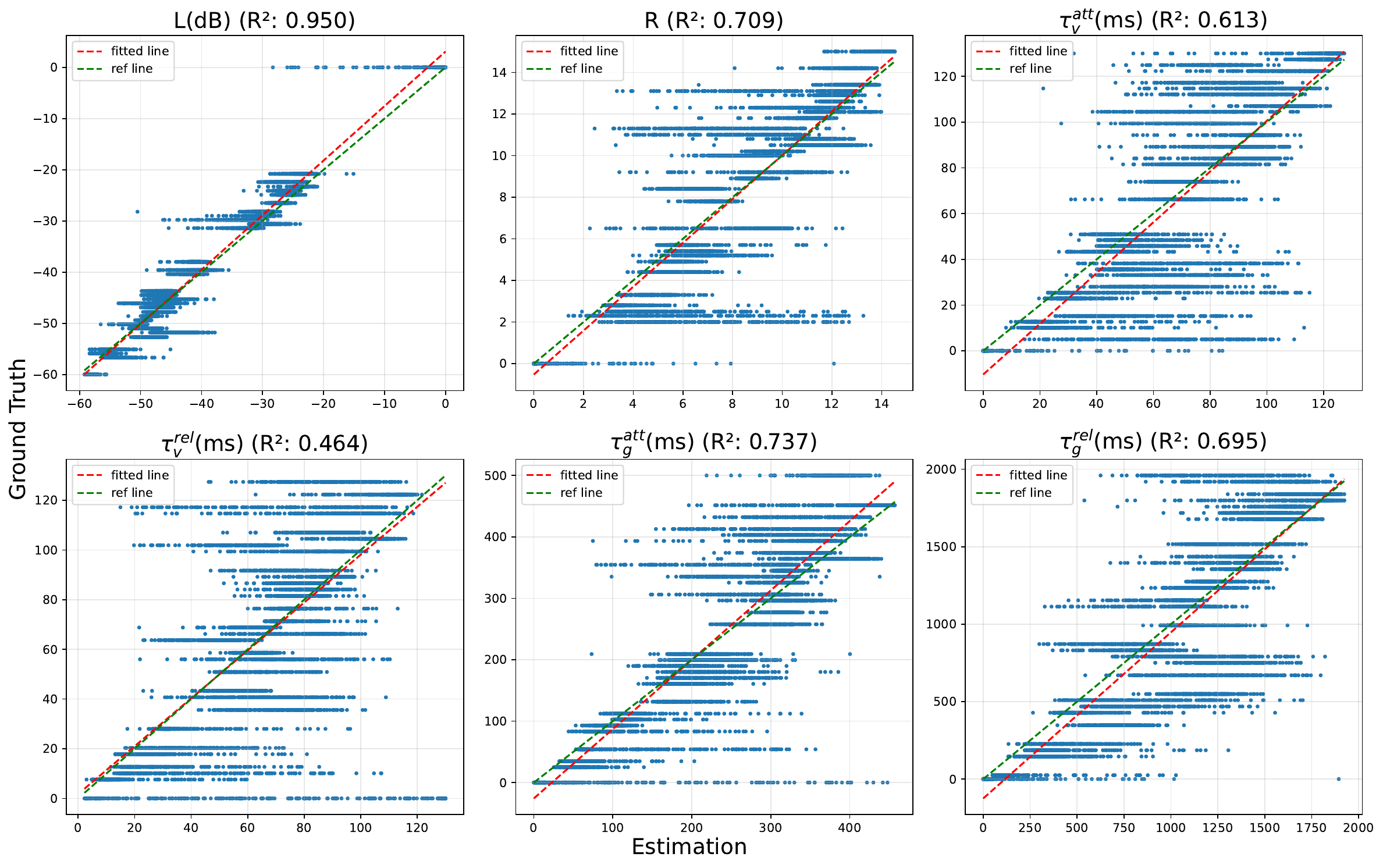}
    \caption{Scatterplot of the real versus estimated values for the 6 DRC parameters using the \ac{mee} model, each plotted with a diagonal reference line (green dashed) and a fitted regression line (red dashed). The coefficient of determination ($R^2$) is shown above each sub-figure, indicating how closely the estimated values align with the true parameters. This figure shows the experimental results using the Med-31 dataset.}
    \label{fig:R2}
\end{figure*}

In addition, we explore the influence of each parameter on the DRC inversion task, specifically, the change in the error of DRC inversion in the case of incorrect estimation. 
We also use the t-SNE method to reduce the dimension and to display the distribution of the estimated parameters in a two-dimensional space as an experimental evaluation. 
In this section, we use the results corresponding to the Med-31 dataset and the experiment with data augmentation and curriculum learning.

Figure~\ref{fig:R2} presents the relationship between the actual (vertical axis) and estimated (horizontal axis) values for the 6 DRC parameters obtained via the \ac{mee} model. Each subfigure corresponds to a different parameter, and each point represents one of the 31 parameter sets used in the experiment.

A diagonal reference line with slope=1 (green dashed line) is included in each plot to indicate perfect agreement between the actual and estimated values. A fitted regression line (red dashed line) is also shown to present any systematic bias in the estimation (i.e., whether the model tends to overestimate or underestimate the parameter). The coefficient of determination ($R^2$) is given above each panel to measure the quality of the regression model.

Overall, the threshold $L$ is the most accurately estimated. One possible reason is that threshold estimation often relies on amplitude-level cues that are comparatively straightforward to extract, making it less sensitive to small errors or noise in the audio or modeling process. 
In contrast, envelope release time $\tau_{v}^{rel}$ can be more difficult to estimate accurately because it depends on the decay characteristics of the signal’s amplitude envelope. These decay regions can be influenced by various factors, such as overlapping transients, signal fluctuations, and the interaction of multiple DRC parameters. Consequently, small inaccuracies in modeling or measurement can compound and lead to larger estimation errors, resulting in a lower $R^2$ value for the envelope release time.
The fitted regression lines (red) show that, for most parameters, the slope is close to but still below 1, indicating a systematic tendency to overestimate or underestimate some parameters. These results suggest that while the \ac{mee}-based approach is promising for some parameters, additional refinement or model calibration may be required to improve estimation accuracy for others.

\begin{figure}[!t]
    \centering
    \includegraphics[width=1.0\linewidth]{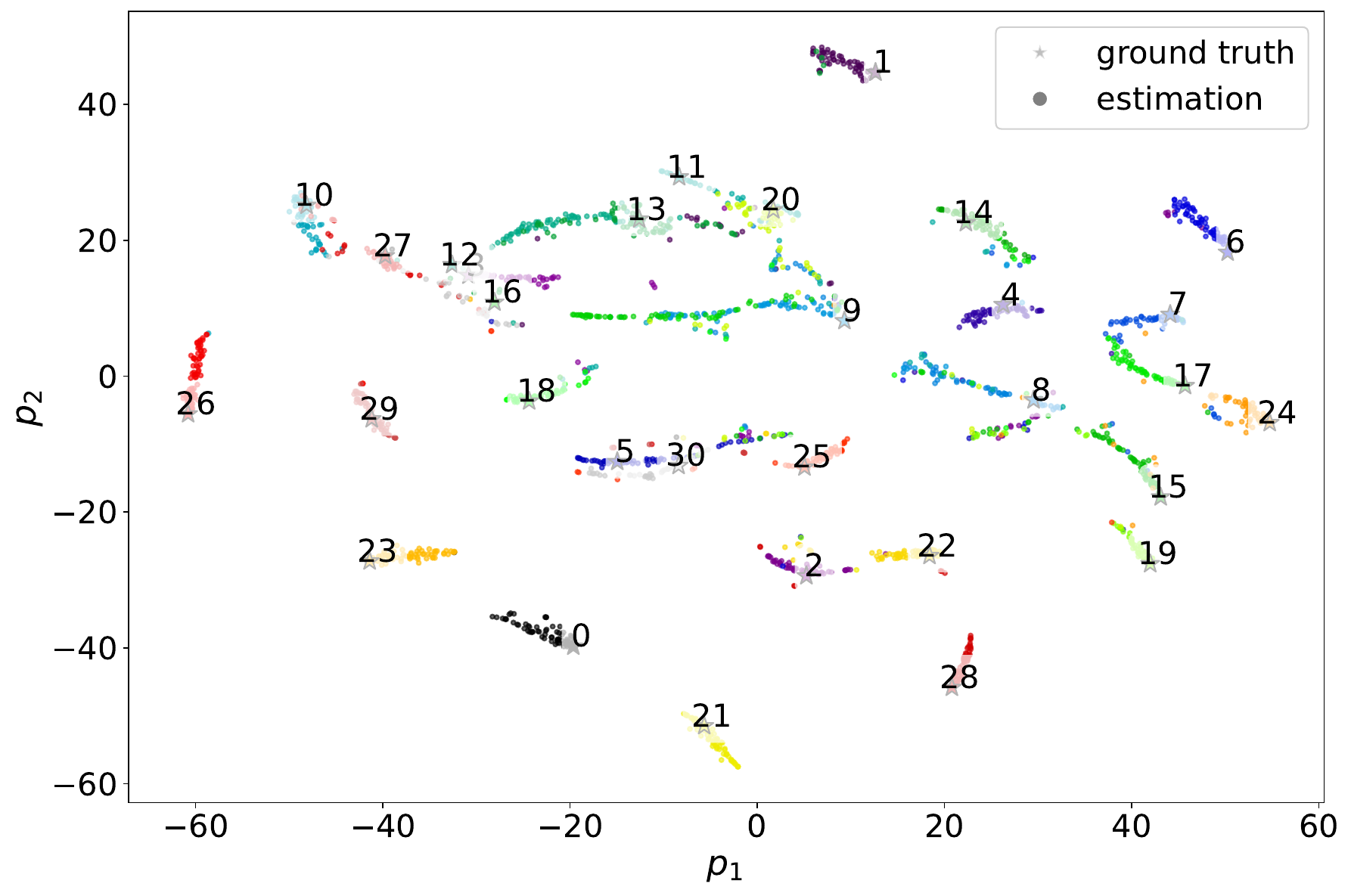}
    \caption{Visualization of the estimated DRC parameters via the \ac{mee} model in two-dimensional space using t-SNE, corresponding to the Med-31 dataset. Each point represents an individual signal, and asterisks denote the actual (reference) parameter values.}
    \label{fig:tsne}
\end{figure}

The t-SNE experiment shows that it is also possible to predict the predefined DRC profile from the estimated parameters provided by the \ac{mee} model.

Figure~\ref{fig:tsne} illustrates the dimensionality reduction results for the 31 distinct \ac{drc} profiles using t-SNE. Each color corresponds to one set of true DRC parameters, and the estimated signals cluster around the asterisk, which marks the true parameter location in the two-dimensional space. Most clusters remain reasonably well separated, indicating that the \ac{mee}-based estimation preserves differences between parameter sets even in higher-complexity scenarios. However, some clusters are more dispersed, reflecting a greater degree of overlap in the two-dimensional embedding and suggesting that some \ac{drc} profiles are more difficult to recognize based solely on the observation of resulting compressed signals.

This experiment is particularly valuable for several reasons. First, dimensionality reduction effectively transforms the complex six-dimensional parameter space into an interpretable two-dimensional visualization, allowing for the verification of parameter estimate consistency and the identification of patterns. Second, the clear clustering features strongly indicate that these profiles can be effectively used for classification tasks, which are particularly useful for automatically identifying or classifying DRC profiles. Finally, the coherent clustering pattern validates the performance of the \ac{mee} model, indicating that it can learn meaningful and consistent parameter representations.

\begin{figure*}[!t]
  \centering
  \subfloat[Results for small datasets.]{\includegraphics[width=1.0\textwidth]{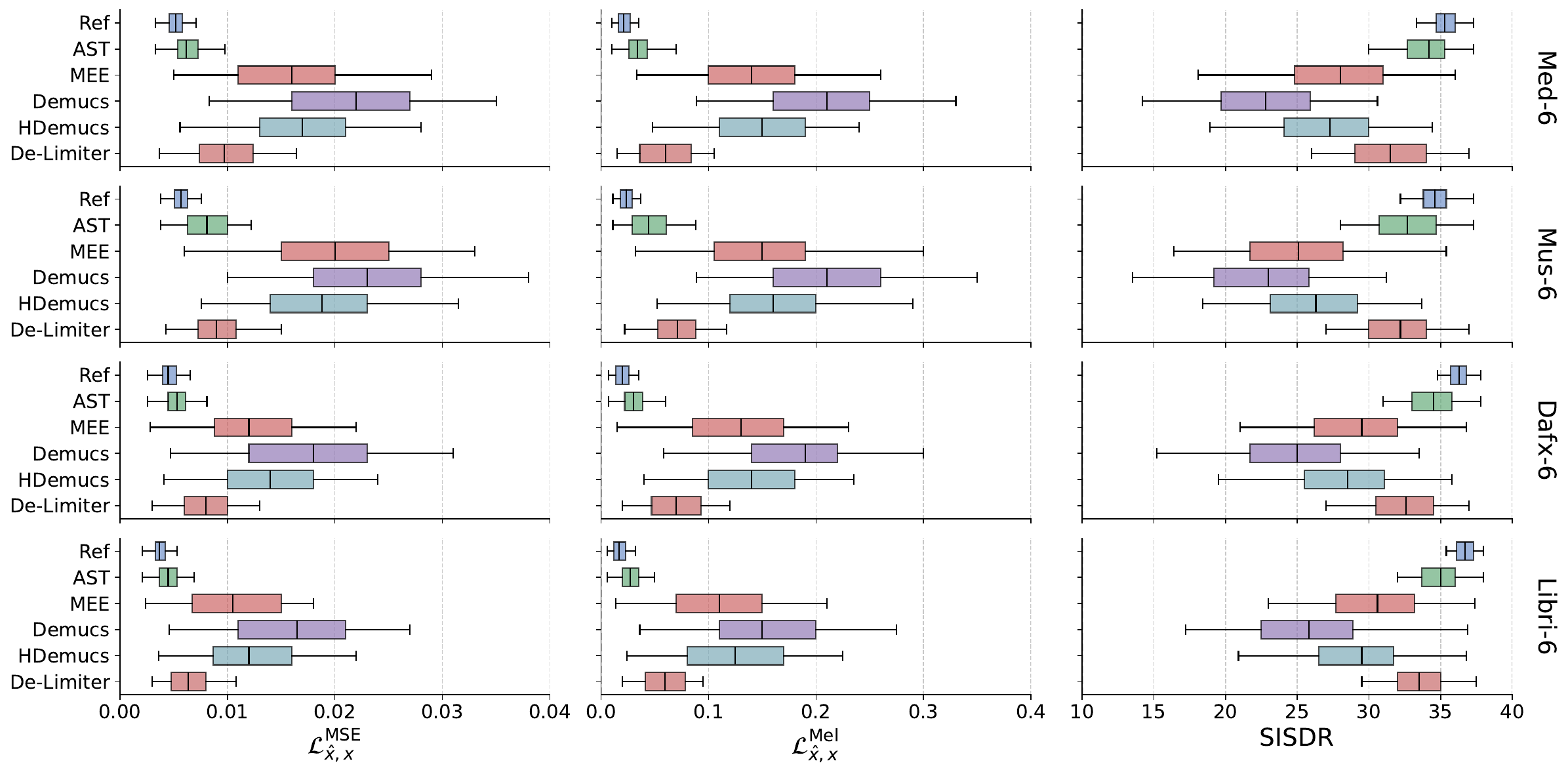} \label{fig:baseline_small}} \\
  \subfloat[Results for large datasets.]{\includegraphics[width=1.0\textwidth]{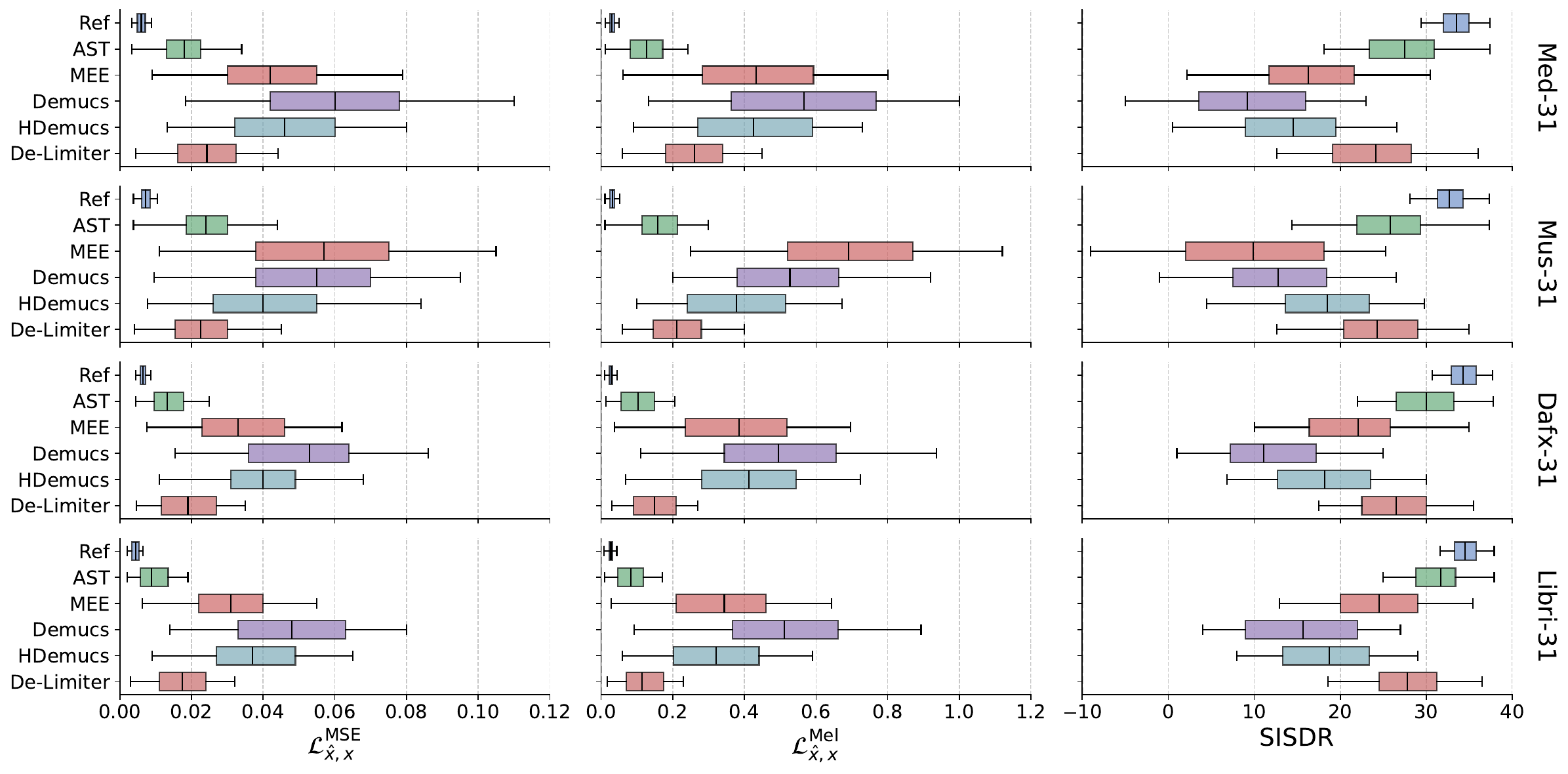} \label{fig:baseline_large}}
  \caption{Box plot comparison of six models across four datasets (rows) and three error metrics (columns) for audio DRC inversion. The proposed \ac{ast} (classification) and \ac{mee} (regression) models are compared against the reference (Ref) and three baseline approaches (Demucs, HDemucs, De-limiter).}
  \label{fig:baseline_score}
\end{figure*}

\subsection{DRC Inversion}

In the end, we conduct a comparative experiment with three state-of-the-art models to further validate the effectiveness of our proposed method.

In \cite{imort2022removing}, the authors proposed to use Demucs \cite{demucs} to remove distortion and clipping applied to guitar tracks for music production and get good results.
Hybrid Demucs (HDemucs) \cite{hdemucs} is the improved version of Demucs. It introduces frequency-domain processing combined with time-domain models to significantly improve audio processing capabilities, especially when processing high-frequency details and complex arrangements.
In \cite{rice2023general}, the authors considered removing multiple audio effects from the same recording and proposed an end-to-end approach that dynamically combines effect-specific removal models. They achieved promising results in \ac{drc} inversion using the HDemucs model.

Music de-limiter network \cite{de-limiter} is a model based on Conv-TasNet that estimates the original signals from heavily compressed signals.
This model estimates the gain parameters of each sample through a neural network and directly performs time-domain element-level multiplication operations on the compressed signal to obtain the original dynamic range of the signal in an end-to-end manner.
This approach is extensible to general DRC inversion tasks, as it inherently models nonlinear gain adjustments, then be used for our work.

Besides, we also use the ground truth parameters $q_{\theta}$ to directly perform the \ac{drc} inversion as a reference for our proposed model, referred to as ``Ref'' in the table. Since the parameters are exact, the resulting inversion error given in this case should be minimal.

For the \ac{ast} and \ac{mee} models, the experimental settings are exactly the same as in the previous subsections. 
For the three baseline models, we use the implementations provided by the original authors and train the models with our datasets and the above-mentioned data augmentation strategy. 
Note that, for the De-Limiter model, for the MedleyDB and Musdb18HQ datasets, we use the weight of a fine-tuned pretrained model, as it is pretrained on the MusdbXL dataset, which is of the same sampling frequency.

Figure~\ref{fig:baseline_score} shows two sets of box plots that summarize the performance of six models across four different datasets for the audio DRC inversion task. Each row corresponds to a particular dataset, while each column presents one of three error metrics: $\mathcal{L}^{\text{MSE}}_{\hat{x},x}$, $\mathcal{L}^{\text{Mel}}_{\hat{x},x}$, and SI-SDR. The top figure reports the results for 6 DRC configurations, whereas the bottom figure corresponds to 31 DRC configurations. In each panel, the models are arranged along the vertical axis, and the distributions reflect performance across multiple audio examples or experimental runs.

In the simplified scenario (Figure~\ref{fig:baseline_small}), the \ac{ast} model shows significant better results: its $\mathcal{L}^{\text{MSE}}_{\hat{x},x}$ and $\mathcal{L}^{\text{Mel}}_{\hat{x},x}$ medians on all the datasets are lower than those of other models, and its SI-SDR median is the highest, indicating that the classification task has high accuracy in most scenarios.
Also, the \ac{ast} model is outstanding in stability (the box plot span is the smallest).

The regression model \ac{mee} has a medium overall performance on the MedleyDB, DAFX, and LibriSpeech datasets. But its minimum $\mathcal{L}^{\text{MSE}}_{\hat{x},x}$ and $\mathcal{L}^{\text{Mel}}_{\hat{x},x}$ values are close to \ac{ast}, and its maximum SI-SDR value is close to Ref, suggesting that there is untapped potential in parameter estimation for the regression task.
Among the baseline models, HDemucs performs better than the pure time domain model Demucs due to the fusion of time-frequency domain feature processing, but are outperformed by the \ac{ast} and \ac{mee} models.

In complex scenarios (Figure~\ref{fig:baseline_large}), the error indicators of all models increased significantly, reflecting the impact of the increase in the number of DRC profiles on the difficulty of the task.
The \ac{ast} model still maintains the best median on all the datasets, but its $\mathcal{L}^{\text{MSE}}_{\hat{x},x}$ and $\mathcal{L}^{\text{Mel}}_{\hat{x},x}$ increase significantly compared with the simplified scenario.
The median performance of the \ac{mee} model deteriorates further, and the variance increases significantly (the span of the box plot increases), highlighting the sensitivity of the regression task to the high-dimensional parameter space.
It is worth noting that the stability of all models decreases with the complexity of the task, but the interquartile range (IQR) of the box plots of \ac{ast} increases the least, indicating that their structural designs are adaptable to complex conditions.

Comprehensive analysis shows that the \ac{ast} model effectively captures the discrete characteristics of the DRC configuration through the classification framework and performs best on all the datasets.
Although the performance of the \ac{mee} model is still insufficient, its performance is better than that of the two baseline models and is close to the De-Limiter model.
At the same time, the experimental results further reveal that time-frequency joint modeling (such as HDemucs) can alleviate the spectral distortion of the pure time domain method (Demucs). The impact of task complexity on the regression model is significantly higher than that on the classification model, which has a guiding significance for model selection in the actual DRC system design.
 

Experimental results show that the performance of the \ac{ast} model is significantly better than that of the \ac{mee} model on the same dataset. We believe that there are two main reasons for this advantage.
First, due to the nature of the task and how the parameters are obtained, the \ac{ast} model performs a classification task, mapping the compressed audio to a finite number of known compressor preset categories. Once the classification is correct, the true parameters corresponding to the preset can be directly referenced for inversion.
In contrast, the \ac{mee} model performs a parameter regression task, and its prediction results must have estimation errors. These parameter errors will be significantly amplified in the subsequent inversion mathematical model, resulting in a decrease in the quality of the original audio signal that is finally reconstructed.
In addition, in audio dynamic range compression, different parameter settings may lead to the same compression effect, which means that there is a "many-to-one" mapping relationship in the parameter space.
This situation is challenging for regression tasks because a unique estimate needs to be found in the parameter space, and this many-to-one relationship makes parameter estimation difficult \cite{come}.
In contrast, the classification task can classify these parameter combinations with the same effect into the same category, thereby avoiding the need for precise estimation in the parameter space.

\section{Conclusion}\label{sec:conc}

In this paper, we have introduced a model-based method coupled with deep learning techniques for \ac{drc} inversion, with a focus on inferring \ac{drc} profiles or \ac{drc} parameters. Our promising results show the effectiveness of the proposed approach for reconstructing audio signals while simultaneously estimating the underlying \ac{drc} profiles. The proposed modified \ac{ast} model obtained better results in \ac{drc} profile type classification and inversion accuracy compared to traditional audio classification methods presented in \cite{fourer2017objective}.


We acknowledge certain limitations and areas for improvement in our method. The model is only evaluated with compression and limiter, but is not evaluated with expander and compander.
Future research directions may involve exploring alternative parameter-based inversion models using neural networks and further numerical experiments considering other types of audio effects.

\end{document}